\documentclass[twocolumn, times]{aastex7}

\usepackage{amsmath}
\usepackage{cases}
\usepackage{graphicx}
\usepackage{subfigure}
\usepackage{natbib}
\usepackage{color}
\usepackage{tabularx}
\usepackage{bm}
\usepackage{threeparttable}
\usepackage{scrextend}

\begin{document}

\title{{\large Dynamical Instability of Multi-planet Systems and Free-floating Planets}}
\author[0009-0004-1650-3494]{Ruocheng Zhai}
\affiliation{Department of Astronomy, Tsinghua University, Beijing 100084, People’s Republic of China}
\affiliation{Department of Astronomy \& Astrophysics, The Pennsylvania State University, University Park, PA 16802, USA}
\email{ruocheng.zhai@gmail.com}

\author[0000-0003-1930-5683]{Man Hoi Lee}
\affiliation{Department of Earth and Planetary Sciences, The University of Hong Kong, Pokfulam Road, Hong Kong, People’s Republic of China}
\affiliation{Department of Physics, The University of Hong Kong, Pokfulam Road, Hong Kong, People’s Republic of China}
\email{mhlee@hku.hk}

\author[0000-0002-4503-9705]{Tianjun Gan}
\affiliation{Department of Astronomy, School of Science, Westlake University, Hangzhou, Zhejiang 310030, People’s Republic of China}
\email{tianjungan@gmail.com}

\author[0000-0001-8317-2788]{Shude Mao}
\affiliation{Department of Astronomy, School of Science, Westlake University, Hangzhou, Zhejiang 310030, People’s Republic of China}
\affiliation{Department of Astronomy, Tsinghua University, Beijing 100084, People’s Republic of China}
\email{shude.mao@westlake.edu.cn}


\begin{abstract}
    The ejection of planets by the instability of planetary systems is a potential source of free-floating planets. We numerically simulate multi-planet systems to study the evolution process, the properties of surviving systems, and the statistics of ejected planets. For systems with only super-Earth planets, we find that the time (in units of the orbital period $P_{1}$ of the innermost planet) for the system to lose the first planet by collision or ejection increases with the semimajor axis of the innermost planet. In contrast, the time (in units of $P_{1}$) for the first close encounter between two planets is identical. These two timescales also depend differently on the orbital spacing between the planets. Most systems with only super-Earths do not have planets ejected. In systems with super-Earths and a cold Jupiter, we discover that a cold Jupiter significantly increases the probability of ejection of the super-Earths by close encounters. Of 38\% of ejected super-Earths, most velocities relative to their parent stars are smaller than $6\ \mathrm{km\ s^{-1}}$. We conservatively estimate that more than 86\% of the surviving two-planet systems in the super-Earths plus cold Jupiter sample are long-term stable by using empirical criteria. Most super-Earths in the remaining two-planet systems are on highly elliptical but stable orbits and have migrated inwards compared with their initial states.
\end{abstract}

\shorttitle{}
\shortauthors{Zhai et al.}

\section{Introduction}\label{sec:intro}
A recent hot topic in planetary science is free-floating planets (FFPs), which bring valuable information about the systems they were born in \citep{Penny2019, Johnson2020, Coleman2024}. They can help to deepen our understanding of the formation and evolution of planets. FFPs are predicted to form in at least two ways: 1) The first is through the direct collapse of gas clouds under self-gravity. Star formation theory points out that when a giant gas cloud is massive enough, the thermal motion of gas molecules cannot balance the gravity of the cloud, and collapse happens. Nuclear reactions are ignited when the pressure and temperature are high enough in the collapsing cloud, and it will form a star or brown dwarf by burning hydrogen or deuterium. However, if the cloud is not massive enough ($\la 13M_\mathrm{J}$, where $M_\mathrm{J}$ is the mass of Jupiter), deuterium burning will not be triggered, and it forms a sub-brown dwarf \citep{Luhman2012}. Sub-brown dwarfs can be seen as FFPs trapped in any gravitational field. 2) The second is by escape from planetary systems formed in protoplanetary disks around young stars (single or binary), in which solid materials merge and grow into planet embryos. Planet embryos can evolve into terrestrial planets and attract some gas to form ice giants. Gas giants are formed when solid cores become massive enough to start runaway gas accretion \citep[e.g.,][]{Pollack1996, Ida2013}. When the planets form on orbits that are too close together, some planets can become dynamically unstable and escape to become FFPs.

These two formation channels of FFPs also determine the two methods to detect them: direct imaging and microlensing. Direct imaging mainly searches for sub-brown dwarfs in the nearby young stellar clusters and star-forming regions \citep{Osorio2000}. Sub-brown dwarfs emit light from their thermal energy, originating from their gravitational potential. By determining their luminosity and temperature, one can constrain their masses. Furthermore, sub-brown dwarfs can be classified by their spectra. A few sub-brown dwarfs of several Jupiter masses have been found of type L  \citep{Best2017}, T \citep{Gagne2015}, and Y \citep{Gagliuffi2020}. In recent observations of the Trapezium Cluster by JWST \citep{Pearson2023}, 9\% of planetary-mass objects are classified as Jupiter-mass binary objects (JuMBOs), which may have formed via direct collapse (however, see  \citealt{Luhman2024} who argued these may be background sources). Since direct imaging requires sources having sufficient brightness to be detected, cooled sub-brown dwarfs and low-mass FFPs ejected from planetary systems cannot be detected. For these cases, microlensing may be the only detection method. Microlensing probes planets through the deflection of light in the gravitational fields of the FFPs, which results in brightness variations of background sources. Thus, it does not require emission from the FFPs, and it analyzes the light curves of background sources. Theoretically, it can detect planets to very low masses (e.g., moons). A dozen or so FFP candidates have been discovered this way with planet masses from sub-Earth mass to several Jupiter masses \citep{Mroz2020, Koshimoto2023}. By analyzing FFPs detected in the Galactic bulge, \citet{Sumi2023} found a large number of FFPs, and the average of their masses falls in the range of super-Earth (see Section \ref{sec:discussion}).

Dynamical analysis is required to study how a planet escapes from a multi-planet system. For two-planet systems, one can use semi-analytic criteria based on the masses and orbital parameters of the planets to decide the stability of the system \citep{Hasegawa1990, Gladman1993, Petrovich2015}. A well-known criterion is that when the orbital separation between two planets on nearly circular orbits is less than $K=2\sqrt{3}$ times the mutual Hill radius (see Section \ref{sec:orbit_parms}), the system can be unstable. However, the complexity of planetary systems with more than two planets requires numerical integration to study their evolution and final state.

Most early works on the instability of multi-planet systems \citep[e.g.,][]{Chambers1996} focused on systems with multiple equal-mass (or similar-mass) planets. They studied the relationship between the time for the first close encounter between two planets (or the time for the first orbit crossing between two planets) and parameters such as the number of planets, planet masses, and their orbital separations. Further studies considered the influence of more complicated orbits (with non-trivial eccentricities, inclinations, etc.) and mass variations \citep[e.g.,][]{Smith2009, Morrison2016, Obertas2017, Rice2023}. They showed that the first close encounter time $t_\mathrm{FCE}$ is tightly correlated with the orbital spacing $K$, satisfying $\log (t_\mathrm{FCE}/P_1) = \alpha K+\beta$, in which $P_1$ is the orbital period of the innermost planet. When the planets are very close to each other ($K<2\sqrt{3}$), the planetary system becomes unstable very quickly. In contrast, when the separations are large ($K>8.4$), the time for the system to become unstable lengthens and deviates from the above relationship \citep{Smith2009}. In addition, two-body mean-motion resonance becomes important when the orbital periods of neighboring planets are near simple integer ratios, and the instability time can either decrease or increase significantly. Progress has also been made in understanding the mechanism of the instability, with \citet{Petit2020} and \citet{Lammers2024} showing that the instability is driven by the overlap of three-body resonances.

To understand the long-term evolution of planet systems and the birth of FFPs, we need to integrate a longer time to simulate the process from the first encounter to the actual loss of any planet, and finally to the stable state of the system. From the simulations, we can learn the properties of the ejected planets and those that remain in the systems. \citet{Matsumoto2017}, \citet{Rice2018}, \citet{Bartram2021}, and \citet{Marzari2025} have performed longer simulations to study the relationship between the time at which the first planet is lost and parameters such as planet masses and orbital separations, eccentricities, and inclinations. However, these studies either did not find any planet ejection due to the parameters adopted for the simulations (see Section \ref{sec:Safronov}) or did not analyze the ejected planets.

Research on the \textit{in-situ} scattering in systems with three warm Jupiter-like planets \citep{Anderson2020, Yuan2024} shows that 15--26\% of Jupiter-like planets are ejected, and orbits of the remaining planets have significant eccentricities. The simulations suggest that FFPs could come from the dynamical evolution of multi-planet systems, and the remaining stable systems have properties similar to those of the observed systems. In addition, \citet{Coleman2024} finds that on average 5 FFPs are produced in circum-binary protoplanetary disk systems, and stellar flybys can perturb two-planet systems, resulting in a 45\% probability to eject planets \citep{Yu2024}. It is also possible for stellar flybys to produce JuMBOS \citep{Pearson2023}, but its formation fraction is smaller than 1\% even under the most optimistic conditions.

Studies on existing planetary systems can also shed light on the origin of FFPs. Using data from radio velocity observations and the \textit{Kepler} mission, \citet{Zhu2018} found that about 30\% of super-Earths coexist with cold Jupiters, and most cold Jupiters coexist with super-Earths \citep[see also][]{Bryan2019, Bryan2024}. This suggests that the formation of super-Earths and cold Jupiters is tightly correlated. Their interactions might produce free-floating super-Earths.

In this work, we numerically simulate planetary systems with only super-Earths and systems with super-Earths and a cold Jupiter. Our study uses more planets and explores a larger range of planetary masses (including a mixture of super-Earths and Jupiters) than previous studies. We analyze unstable events in these systems (in particular, the interaction between giant planets and super-Earths) and the products from their evolution, including ejected planets and surviving planetary systems. In addition, we explore the influences of orbital spacings and distances between planets and their host stars on the evolution of planetary systems and the probability of ejection. In Section \ref{sec:method}, we introduce the parameters of the simulated planetary systems and the numerical integration methods. We show the results of super-Earths only and super-Earths-cold-Jupiter simulations in Section \ref{sec:simulation}. In Section \ref{sec:discussion}, we discuss the influence of orbital spacing $K$ and Safronov number and the stability of surviving two-planet systems in the super-Earth-cold-Jupiter sample. We conclude our work in Section \ref{sec:conclusion}.

\section{Orbital Parameters and Simulation Methods}\label{sec:method}
\subsection{Orbital Spacing of Planets}\label{sec:orbit_parms}

We begin our simulations after the planets have formed and the protoplanetary gas disk has dispersed.
We study planetary systems with multiple planets orbiting a host star, where the host has one solar mass $M_\star = 1M_\odot$ and one solar radius $R_\star = 1R_\odot$. (Systems with different host star mass can be transformed to systems with $M_\star = 1M_\odot$.) The mass and radius of the $i$-th planet ordered from the inside to the outside are $m_i$ and $r_i$, respectively. Orbital parameters of planets include the semimajor axis $a$, the eccentricity $e$, the inclination $i$, the longitude of the ascending node $\Omega$, the argument of periapsis $\omega$, and the mean anomaly $M$.

In multi-planet systems, we scale how close two orbits are with the mutual Hill radius \citep{Chambers1996}:
\begin{equation}\label{eq:Hill_radius}
    R_{\mathrm{H}_{i,i+1}} = \left(\frac{ m_i+m_{i+1}}{3M_*}\right)^{1/3}\frac{a_i+a_{i+1}}{2}.
\end{equation}
The separation between two neighboring planets is: 
\begin{equation}\label{eq:separation}
    a_{i+1} - a_i = KR_{\mathrm{H}_{i,i+1}}.
\end{equation}
A smaller orbital spacing $K$ means that the orbits of two planets are closer and interact more strongly. As a result, the system is more likely to become unstable.

\subsection{Numerical Integration Methods}\label{sec:int_method}

We use the $N$-body integration package \texttt{REBOUND} \citep{Rein2012} to calculate the evolution of multi-planet systems. We use two integrators under different conditions: IAS15 \citep{IAS15} and Mercurius \citep{Mercurius}. IAS15 is a high-order and non-symplectic integrator that adaptively adjusts the time step according to the separations between the particles in a system. Although it is not symplectic, it can still guarantee energy conservation by suppressing the integration error to machine precision. Mercurius is a hybrid symplectic integrator that combines IAS15 and WHFAST \citep{WHFAST, WHFAST_Rein}, in which WHFAST is a second-order symplectic integrator that can integrate fast and accurately when particles are far away from each other. Mercurius switches to IAS15 when distances between particles are small and back to WHFAST when they are large, at a threshold of several times the mutual Hill radius, to handle close encounters.

Since general relativistic (GR) apsidal precession can affect the stability of orbits close to the host star, we introduce GR corrections to the numerical integration. We use \texttt{gr-potential} \citep{Nobili1986} in \texttt{REBOUNDx} \citep{Tamayo2020} as a correction term. This algorithm applies to systems dominated by a central massive object and keeps WHFAST symplectic because its potential simply gives an additional kick to the linear momentum of each particle.

The simulations will have three types of events: 1. Planet-planet collision (pp) or planet-star collision (ps), which happens when the sum of the radii of two particles is less than the distance between them. When two particles collide, we merge them, keeping the total mass and volume the same and momentum conserved; 2. Planet ejection (ej), which we define to occur when the distance of a planet from the center of mass of the system is larger than 1000 au. The ejected planet is removed from the system; 3. Close encounter between the planets, which is the situation when the distance between two planets is smaller than their mutual Hill radius (but larger than the sum of their physical radii).

\subsection{Numerical Integration Settings}

For the initial conditions of our planetary systems, we first set the semimajor axis of the innermost orbit $a_1$. With a fixed orbital spacing $K$, we obtain semimajor axes of other planets by satisfying Equations (\ref{eq:Hill_radius}) and (\ref{eq:separation}) simultaneously. This scaling of planetary separations helps to compare with previous works that adopted similar settings. Following \citet{Anderson2020}, we randomly choose other parameters: $e$ in [0.01, 0.05], $i$ in $[0^\circ, 2^\circ]$, $\Omega$, $\omega$ and $M$ in $[0^\circ, 360^\circ]$. The small but non-zero ranges for the eccentricities and inclinations take into account the excitation and damping of eccentricities and inclinations during planet formation, including the damping by planet-disk interactions \citep{Kley2012}. 

Considering that our samples might be unstable on a wide range of timescales \citep{Chambers1996, Rice2023}, we divide the numerical integration into two phases \citep{Anderson2020, Yuan2024}. In Phase 1 which is likely to be more chaotic, we integrate with the IAS15 integrator to guarantee accuracy. The total integration time of Phase 1 is $10^{6} P_1$; the integration timestep is $10^{-3} P_1$, where $P_1$ is the initial period of the innermost orbit. In Phase 2, we integrate all systems with 2 or more planets remaining after Phase 1. Since this phase requires much longer integration to ensure all systems end in stable states, we use the Mercurius integrator. The total integration time of Phase 2 is $10^{8} P_1^\prime$, with a timestep of $10^{-2} P_1^\prime$, where $P_1^\prime$ is the period of the innermost orbit at the beginning of Phase 2. The switching threshold between IAS15 and WHFAST of Mercurius is 3 times the Hill radius. In the next section, we will introduce the settings on the planet mass, radius, innermost orbit, and orbital spacing $K$.

\section{Numerical Simulations}\label{sec:simulation}
\subsection{Super-Earths Only Systems}\label{sec:superearth}

To study the interaction between cold Jupiters and super-Earths, we first simulate super-Earths only systems to discover their evolution as a prelude. Most previous studies on the instability of multiple equal-mass planetary systems fix the semimajor axis of the innermost orbit and change the mass and orbital spacing $K$ of the planets, and they focus on the first close encounter time \citep[e.g.,][]{Chambers1996, Smith2009, Rice2023}. Here, we attempt to study systems at various distances from the host star by changing $a_1$. The initial conditions as listed in Table \ref{tab:superearth_initial} are: 5 super-Earths, each with mass $5M_\oplus$ and radius $5^{1/3}R_\oplus$; The semimajor axis of the innermost orbit $a_1$ is randomly chosen from a Gaussian distribution governed by $a_{1,0} \pm 0.01\mathrm{au}$, and the orbital spacing is $K=5$. Here $a_{1,0}$ determines the initial semimajor axes of all systems in a sample. We simulate 180 systems for each $a_{1,0}$.

\begin{table}[tb]
    \centering
    \caption{Initial Conditions of Super-Earths Only Simulations}
    \label{tab:superearth_initial}
    \begin{tabular}{l c r}
        \hline
        \hline
        Number of planets & \quad & 5 super-Earths \\
        Planet mass $m_\mathrm{p}$ & \quad & 5$M_\oplus$ \\
        Planet radius $r_\mathrm{p}$ & \quad & $5^{1/3}R_\oplus$ \\
        $a_1$ & \quad & Gaussian distribution of $a_{1,0} \pm 0.01\ \mathrm{au}$ \\
        $K$ & \quad & 5 \\
        \hline
    \end{tabular}
\end{table}

\begin{figure}[tb]
    \centering
    \includegraphics[width=1\linewidth]{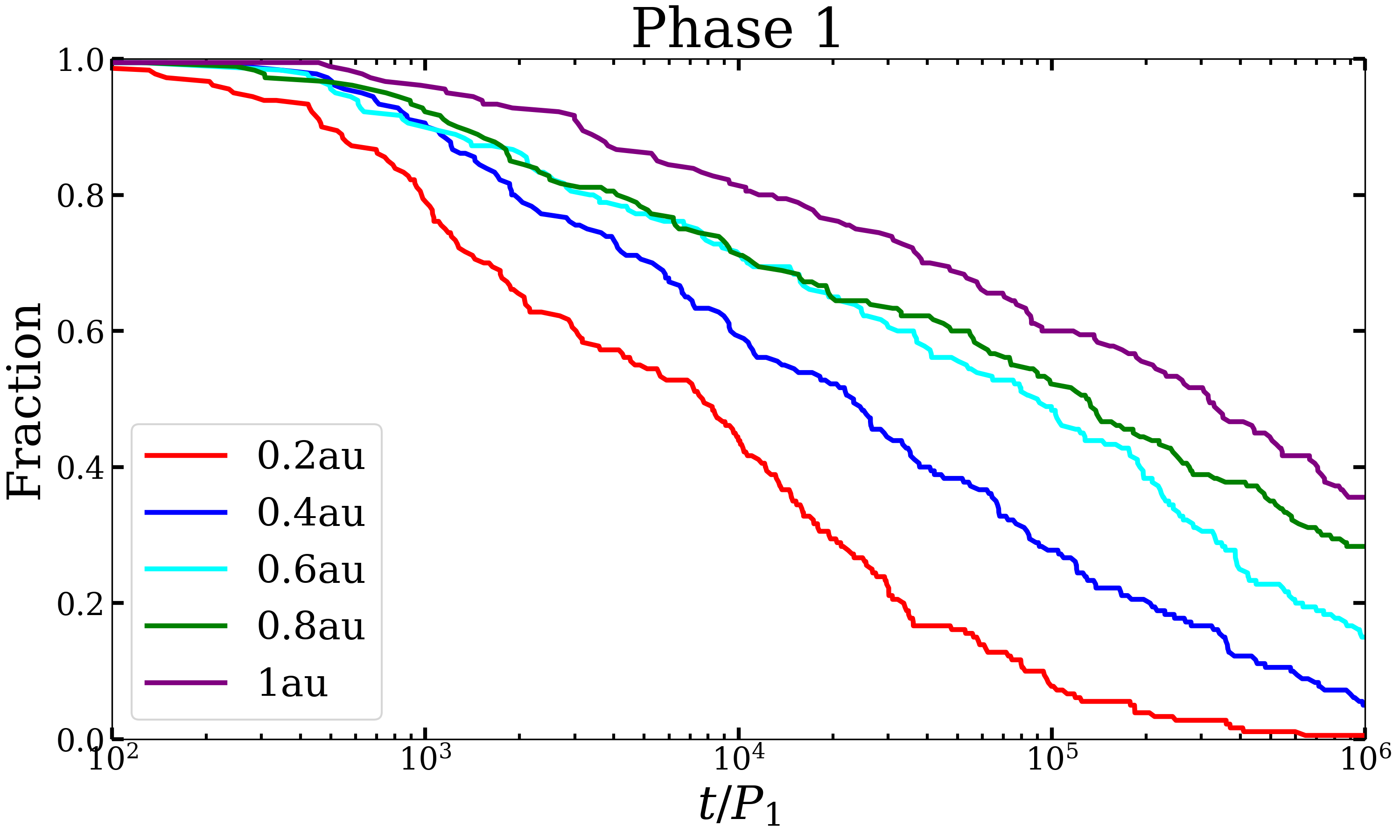}
    \caption{The fraction of five-planet systems in the super-Earths only simulations with $K = 5$ as a function of time, which is normalized by the period $P_{1}$ of the initial innermost orbit of each system. Five curves are shown for $a_{1,0} = 0.2,\ 0.4,\ 0.6,\ 0.8$, and $1$ au, where $a_{1,0}$ defines the initial distribution of the semimajor axis of the innermost orbit (see Table \ref{tab:superearth_initial}). In samples with smaller $a_{1,0}$, the fraction of five-planet systems decreases faster with time.}
    \label{fig:superearth_compare_a}
\end{figure}

It takes longer for systems with larger $a_1$ to become unstable. In Figure \ref{fig:superearth_compare_a}, we show curves of the fraction of planetary systems that have not experienced the first unstable event (i.e, a collision or ejection that changes the number of planets from 5 to 4) changing with the time normalized by the period $P_{1}$ of the initial innermost orbit of each system. We see that as the planets become farther away from the host, the time when systems become unstable grows accordingly.

\begin{figure}[tb]
    \centering
    \includegraphics[width=1\linewidth]{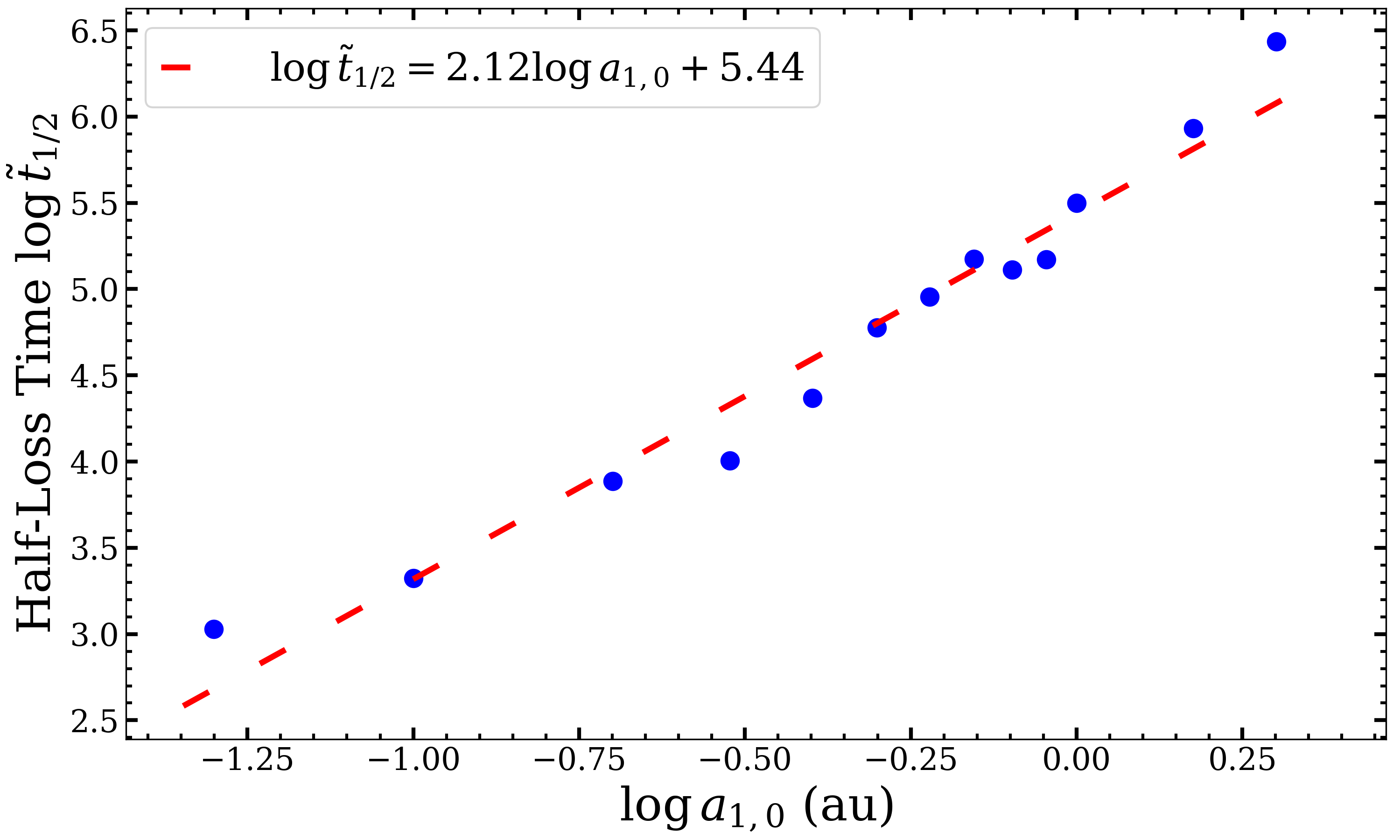}
    \caption{Half-loss time vs. the semimajor axis of the innermost orbit for the super-Earths only samples with $K = 5$. The red dashed line shows the linear regression result. $\tilde{t}$ is the time normalized by $P_1$: $\tilde{t} = t/P_1$.}
    \label{fig:superearth_half_t}
\end{figure}

To compare the speed at which samples of different $a_{1,0}$ become unstable, we define a half-loss time $\tilde{t}_{1/2}$ ($\tilde{t} = t/P_1$) as the time in units of $P_{1}$ when half of the systems in a sample have experienced the first unstable event and lost a planet. This is when the fraction of five-planet systems has decreased to 50\% (see Figure \ref{fig:superearth_compare_a}). As Figure \ref{fig:superearth_half_t} shows, in logarithm coordinates, the half-loss time of the super-Earths only samples increases with the distance from the host. By linear regression:
\begin{equation}
    \log \tilde{t}_{1/2} = b \log a_{1,0} + c,
\end{equation}
we have $b = 2.12 \pm 0.13$ and $c = 5.44 \pm 0.07$.

\begin{figure}[tb]
    \centering
    \includegraphics[width=1\linewidth]{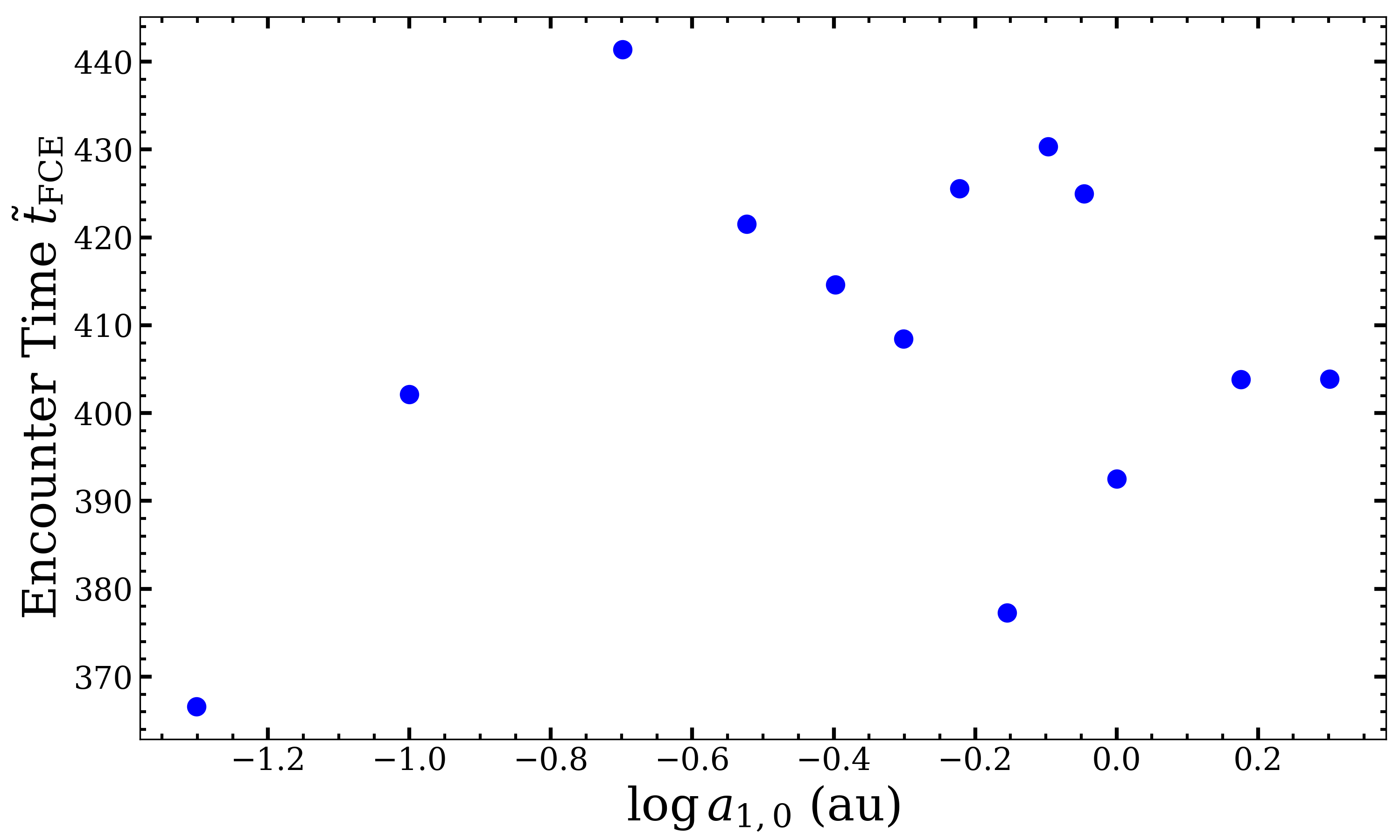}
    \caption{The first close encounter time ($\tilde{t}_\mathrm{FCE}$) vs. the initial semimajor axis of the innermost orbit $a_{1,0}$ for super-Earths only samples with $K=5$. The first close encounter time depends little on the distance between the planets and the host.}
    \label{fig:superearth_encounter_t}
\end{figure}

We calculate half-loss time $\tilde{t}_{1/2}$ as a statistical quantity, while \citet{Chambers1996} and others considered the first close encounter time when they studied the instability of multi-planet systems. So we also calculate the median of the first close encounter time in units of $P_{1}$, $\tilde{t}_\mathrm{FCE}$, in each sample as another viewpoint. In Figure \ref{fig:superearth_encounter_t}, we show $\tilde{t}_\mathrm{FCE}$ for different $\log a_{1,0}$. Interestingly, there is no clear dependence of the first close encounter time on the distance between the planets and the host. So even though many previous works \citep{Chambers1996, Smith2009, Rice2023} fixed $a_{1,0}$ when numerically integrating at various orbital spacing $K$, their results should still represent the statistical properties of close encounters for samples of different innermost orbits. However, the strikingly different behaviors of $\tilde{t}_{1/2}$ and $\tilde{t}_\mathrm{FCE}$ mean that there is a complex dynamical process between the first close encounter and the first planet loss in the planetary systems. Since this process depends significantly on the scale of the systems, we cannot simply regard the first close encounter as the measure of planetary systems becoming unstable.

\begin{figure}[tb]
    \centering
    \includegraphics[width=1\linewidth]{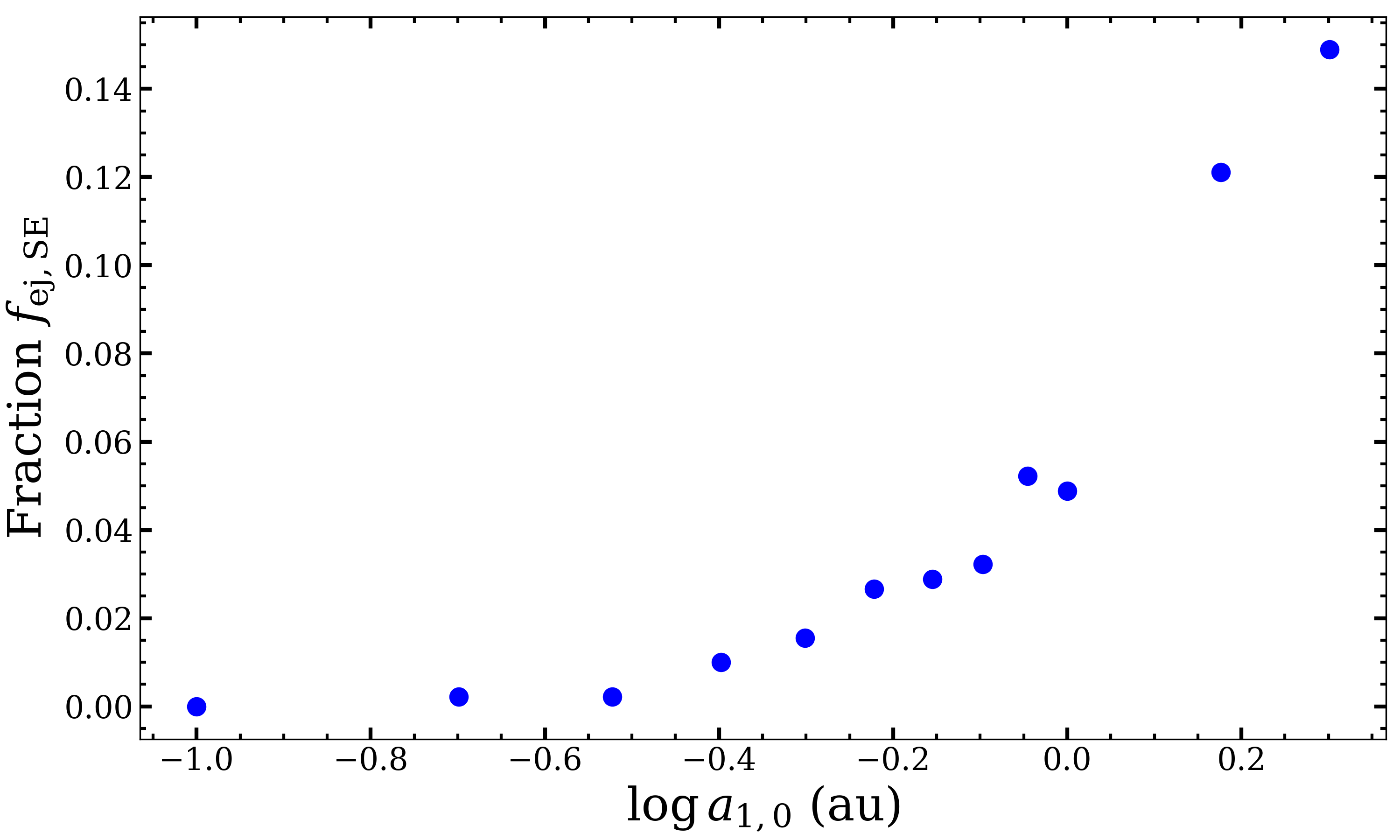}
    \caption{The fraction $f_{\mathrm{ej,SE}}$ of all super-Earths that have been ejected vs.\ the initial semimajor axis of the innermost orbit $a_{1,0}$ for super-Earths only samples with $K = 5$.}
    \label{fig:superearth_ejection_fraction}
\end{figure}

In Figure \ref{fig:superearth_ejection_fraction}, we show the fraction $f_{\mathrm{ej,SE}}$ of all super-Earths that have been ejected by the end of Phase 2 for different $\log a_{1,0}$. It is $< 2\%$ for $a_{1,0} \le 0.5$ au and reaches $15\%$ for $a_{1,0} = 2$ au. The increase of $f_{\mathrm{ej,SE}}$ with the distance from the host can be understood in terms of the Safronov number (see Section \ref{sec:Safronov}).

\citet{Marzari2025} has recently performed simulations with $a_{1,0} = 5\,$au and stated that both the time of first close encounter and the time of first planet loss should scale with $P_{1}$.
As we have just shown, this scaling is correct only for the time of first close encounter.
\citet{Rice2018} have studied the effects of changing $a_{1,0}$ using simulations with four planets of mass $m = 1 \times 10^{-5} M_\sun$ on nearly circular and coplanar orbits with $K = 5$ and $a_{1,0} = 0.01$, $0.1$, $1$, $10$, and $100\,$au.
They found that the distribution of the first close encounter time is nearly independent of $a_{1,0}$, which is in agreement with our Figure \ref{fig:superearth_encounter_t}.
In addition, they found that the distribution of the time for the loss of the first planet is nearly identical to the distribution of encounter time at $a_{1,0} = 0.01\,$au, whereas the time for the loss of the first planet is much longer than the encounter time for the majority of the systems at $a_{1,0} = 100\,$au.
This is also consistent with our Figures \ref{fig:superearth_half_t} and \ref{fig:superearth_encounter_t}.
By using $\tilde{t}_{1/2}$ to characterize the distribution of the time of first planet loss, we are able to quantify the dependence of $\tilde{t}_{1/2}$ on $a_{1,0}$, as shown in Figure \ref{fig:superearth_half_t} and Equation (\ref{eq:half_loss_t}) (see Section \ref{sec:Spacing} for further discussion).

\subsection{Super-Earths plus cold-Jupiter Systems}\label{sec:SECJ}

As we mentioned in Section \ref{sec:intro}, \citet{Zhu2018} found that 30\% of super-Earths co-exist with cold Jupiters, and almost all cold Jupiters co-exist with super-Earths. This implies that the evolution of super-Earths and cold Jupiters may be highly correlated. Thus, we numerically simulate planetary systems with both super-Earths and cold Jupiters. Based on Section \ref{sec:superearth}, we add a Jupiter at the outermost of the super-Earth systems, whose orbit is also determined by Equation (\ref{eq:separation}). We focus on the case with $a_{1,0} = 0.5$ au and $K = 5$ (see Table \ref{tab:SEJ_initial} for the full set of initial conditions). These systems should undergo more chaotic evolution with the close association of the super-Earths and the cold Jupiter. By comparing with the super-Earths only samples, we can study the influence of the cold Jupiter on the super-Earths. We simulate 180 systems, identical to the number of pure super-Earth systems we studied.

\begin{table}[tb]
    \centering
    \caption{Initial Conditions of Super-Earths-cold-Jupiter Simulations}
    \label{tab:SEJ_initial}
    \begin{tabular}{l c r}
        \hline
        \hline
        Number of planets & \quad & 5 super-Earths \& 1 cold Jupiter \\
        Planet mass $m_\mathrm{p}$ & \quad & 5$M_\oplus, M_\mathrm{J}$ \\
        Planet radius $r_\mathrm{p}$ & \quad & $5^{1/3}R_\oplus, R_\mathrm{J}$ \\
        $a_1$ & \quad & Gaussian distribution of $0.5 \pm 0.01\ \mathrm{au}$ \\
        $K$ & \quad & 5 \\
        \hline
    \end{tabular}
\tablecomments{$M_\mathrm{J}$ and $R_\mathrm{J}$ are the mass and radius of Jupiter, respectively.}
\end{table}

\begin{figure*}[tb]
    \centering
    \includegraphics[width=0.44\linewidth]{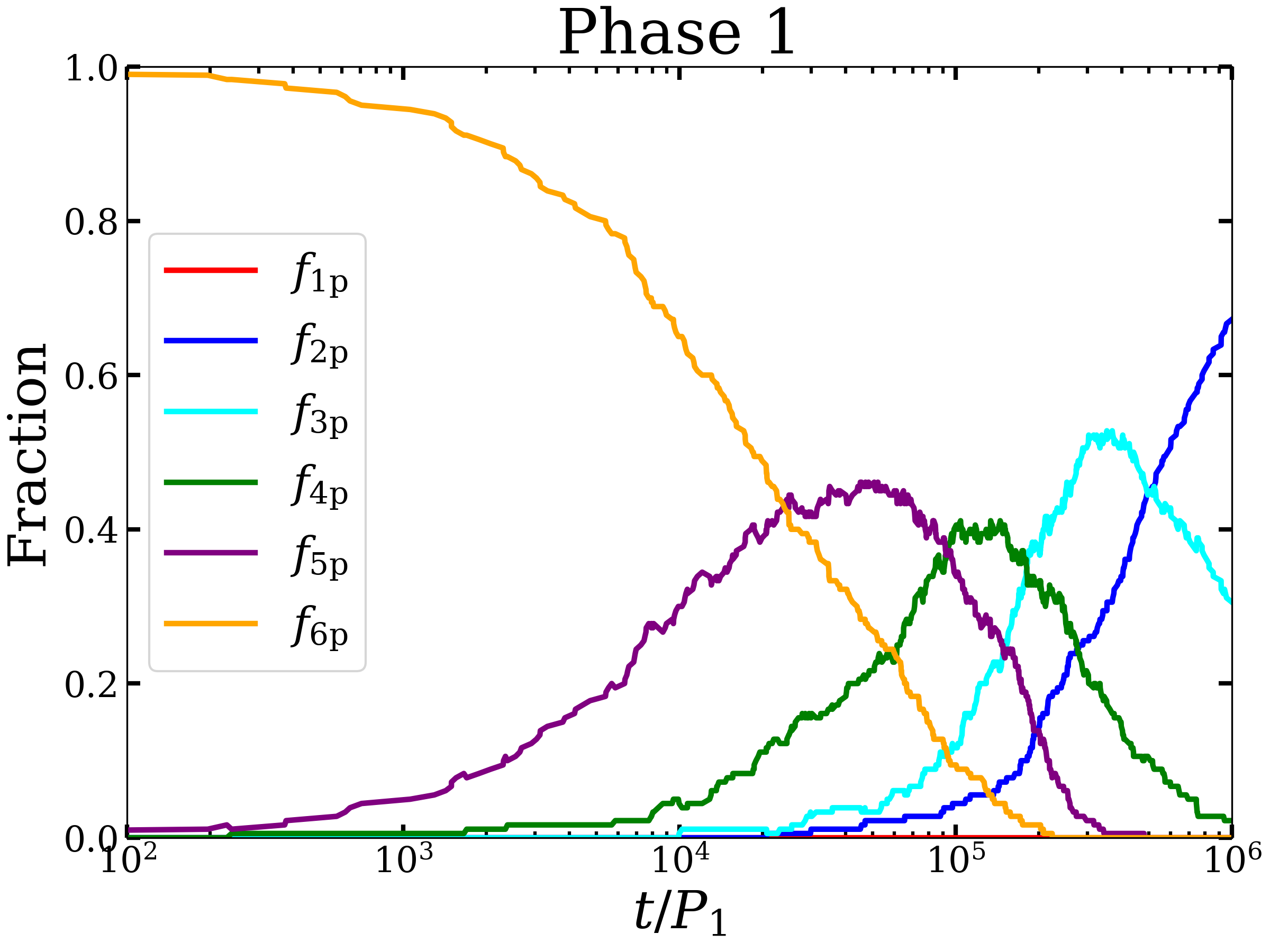}
    \includegraphics[width=0.44\linewidth]{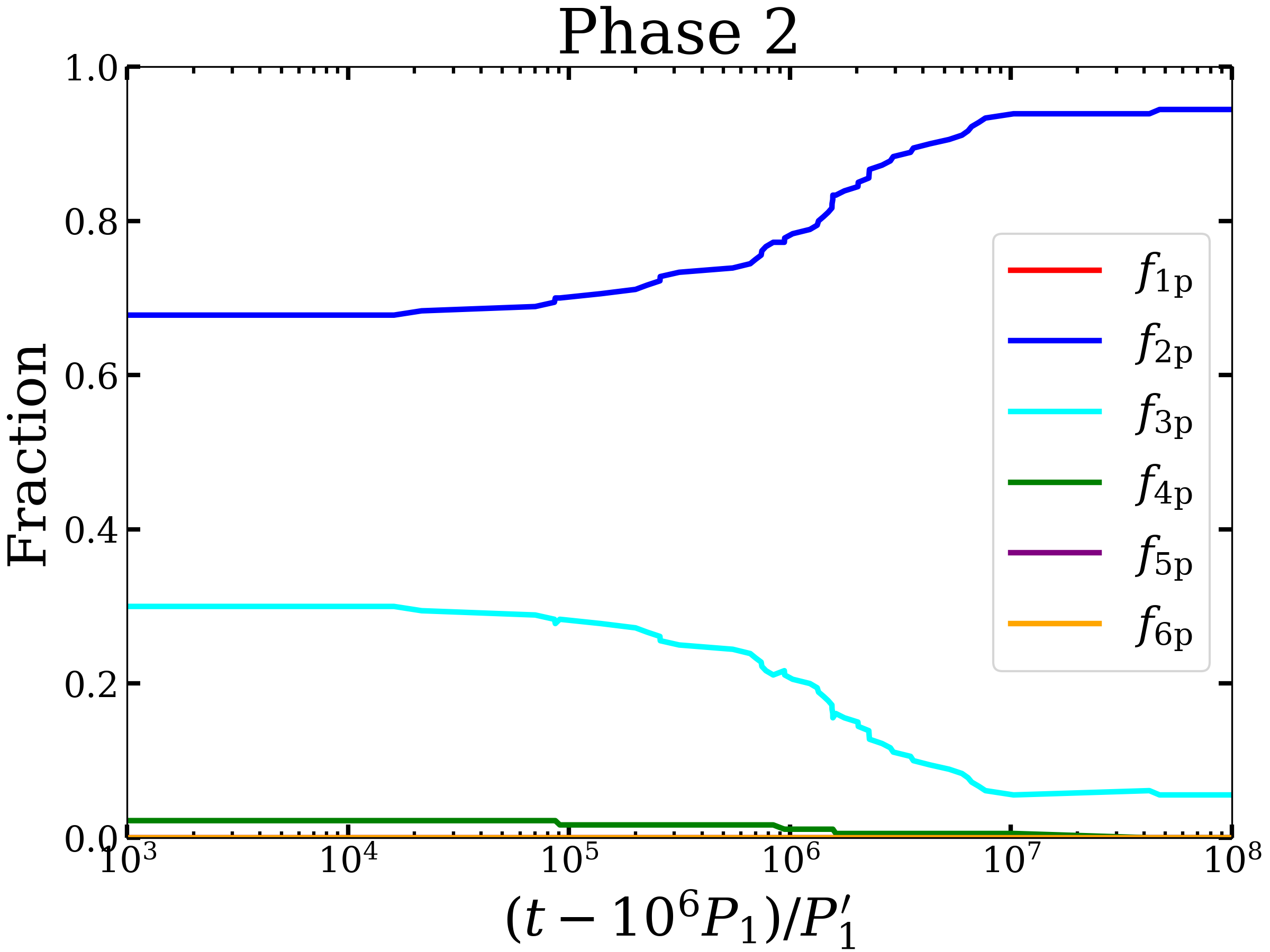}
    \caption{Evolution of the fractions of planetary systems with different numbers of planets in the super-Earth-cold-Jupiter sample with $a_{1,0} = 0.5$ au and $K =5$. Phases 1 and 2 are in the left and right panels, respectively. Curves labeled $f_{n\mathrm{p}}$ are the fractions of systems with $n$ planets. $P_1$ and $P_1^\prime$ are the periods of the innermost orbits at the beginning of Phase 1 and Phase 2, respectively.}
    \label{fig:SEJ_frac}
\end{figure*}

In Figure \ref{fig:SEJ_frac}, we show the evolution of the fractions of planetary systems with different numbers of planets in the super-Earth-cold-Jupiter sample in Phases 1 and 2. In Phase 1, all systems evolve fast, most of which become three- or two-planet systems. In Phase 2, some three-planet systems become unstable, producing more two-planet systems. No single-planet system appears in this simulation. 94\% of planetary systems end up with 2 planets, and 6\% end up with 3 planets (see $K = 5$ in Table \ref{tab:SEJ_K_stats} for more information).

\begin{table}[tb]
    \centering
    \caption{Statistical Outcomes of Super-Earths-cold-Jupiter Samples with different orbital spacing $K$}
    \label{tab:SEJ_K_stats}
    \begin{tabular}{c c c c c c c c}
        \hline
        \hline
        $K$ & $f_\mathrm{1p}$ & $f_\mathrm{2p}$ & $f_\mathrm{3p}$ & $f_\mathrm{ej,SE}$ & $e_\mathrm{f, med}$ & $e_\mathrm{f, 10}$ & $e_\mathrm{f, 90}$ \\
        \hline
        4 & 0\% & 96\% & 4\% & 38\% & 0.23 & 0.09 & 0.45 \\
        5 & 0\% & 94\% & 6\% & 38\% & 0.21 & 0.08 & 0.44 \\
        6 & 1\% & 93\% & 7\% & 38\% & 0.22 & 0.07 & 0.45 \\
        \hline
    \end{tabular}
    \tablecomments{$f_{n\mathrm{p}}$ is the fraction of systems with $n$ planets remaining. $f_\mathrm{ej,SE}$ is the fraction of all super-Earths that have been ejected. $e_\mathrm{f, med}$, $e_\mathrm{f, 10}$, and $e_\mathrm{f, 90}$ are the median, 10th percentile, and 90th percentile values of the eccentricity of the super-Earths orbit in surviving two-planet systems.}
\end{table}

In the super-Earth-cold-Jupiter sample, a significant number of ejected planets appear. All of them are super-Earths or new planets born from collisions between super-Earths. It shows that the super-Earths cannot effectively perturb the orbit of the cold Jupiter, which is expected from their large mass difference. 38\% of all super-Earths are ejected, with 91\% of them escaping in Phase 1. In contrast, the super-Earths only sample with $a_{1,0}=0.5\ \mathrm{au}$ has no ejection in Phase 1 and $< 2\%$ ejection in Phase 2. This indicates that the cold Jupiter significantly increases the probability of super-Earths ejecting. Because of the large mass of the cold Jupiter, super-Earths having close encounters with it have a high probability of being ejected. 

\begin{figure*}[tb]
    \centering
    \includegraphics[width=0.45\linewidth]{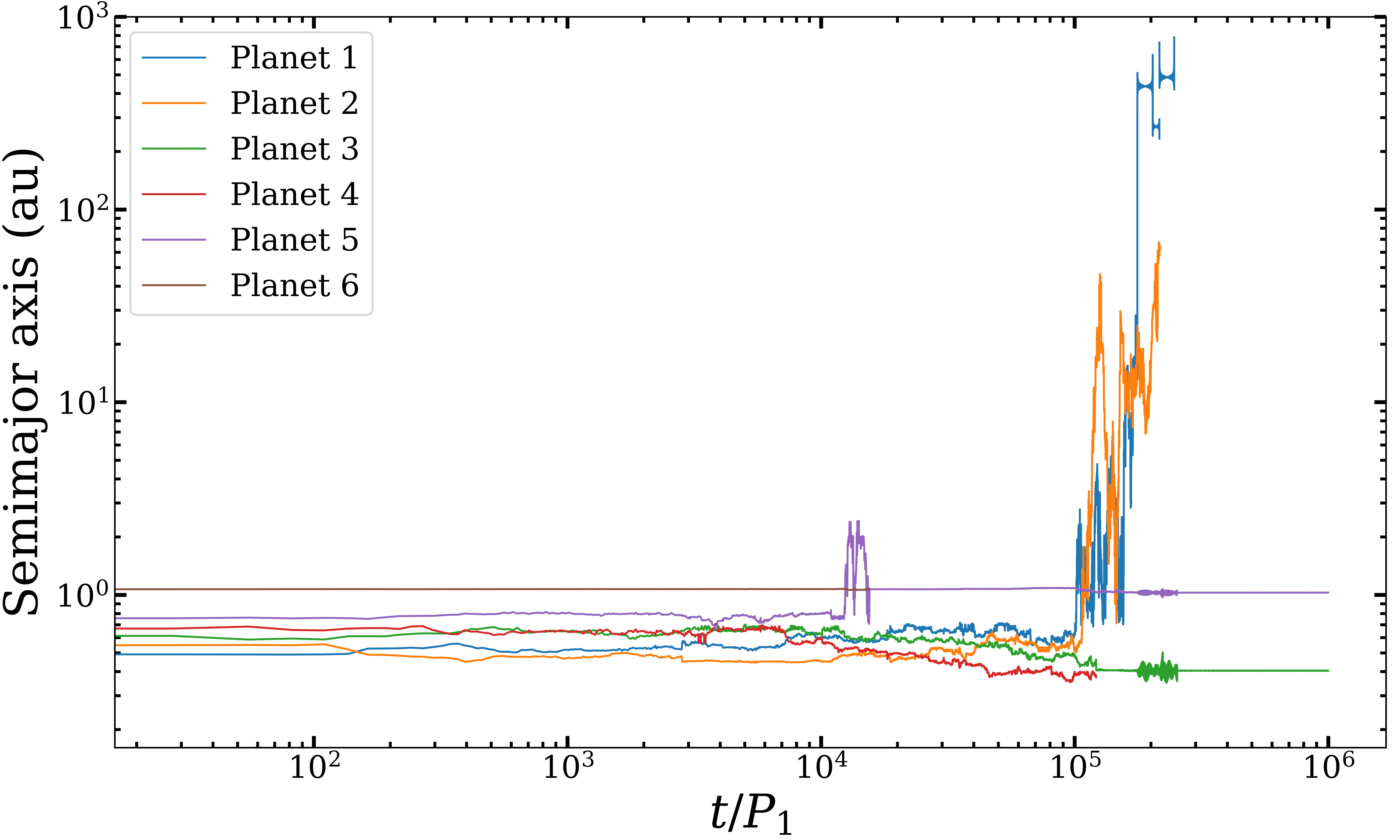}
    \includegraphics[width=0.45\linewidth]{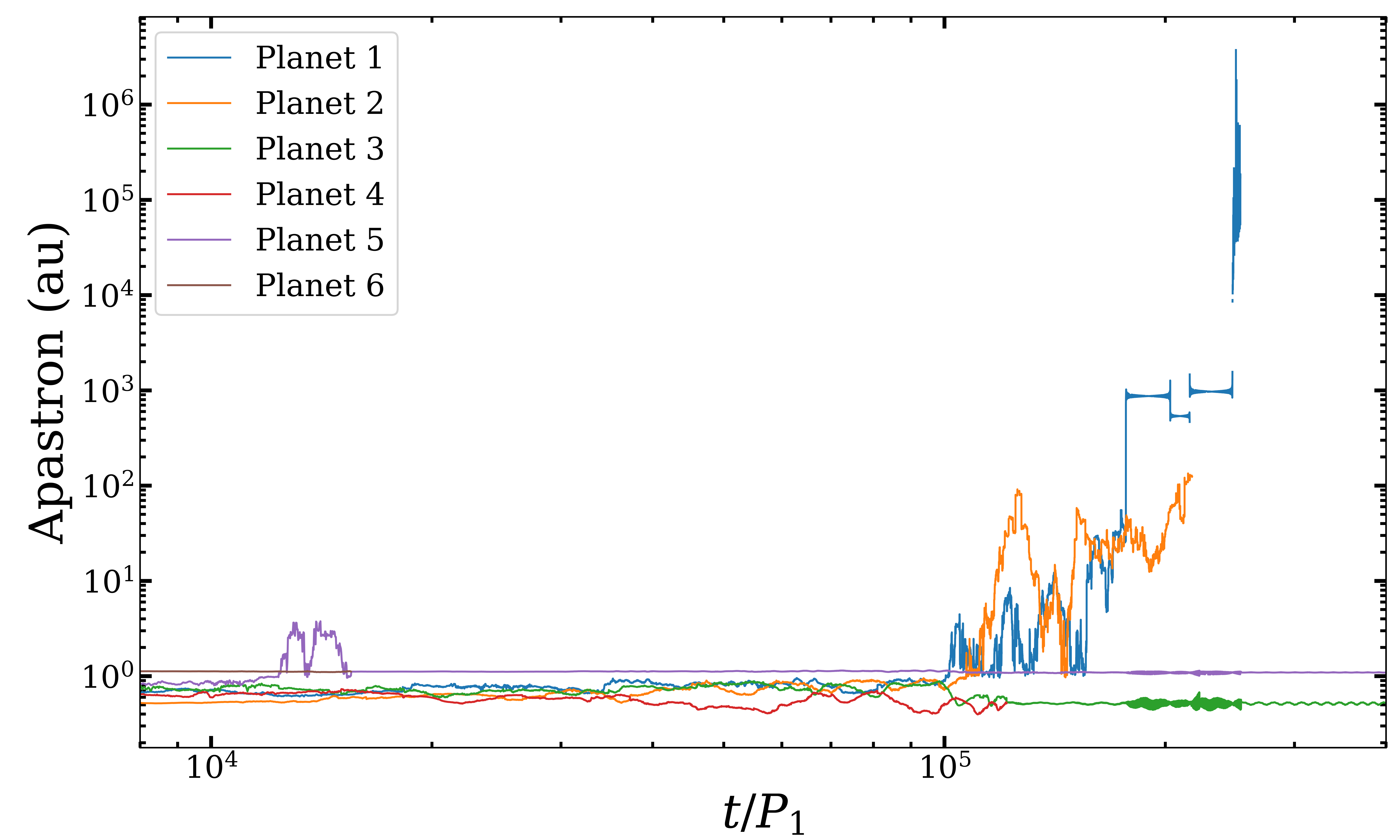}
    \includegraphics[width=0.45\linewidth]{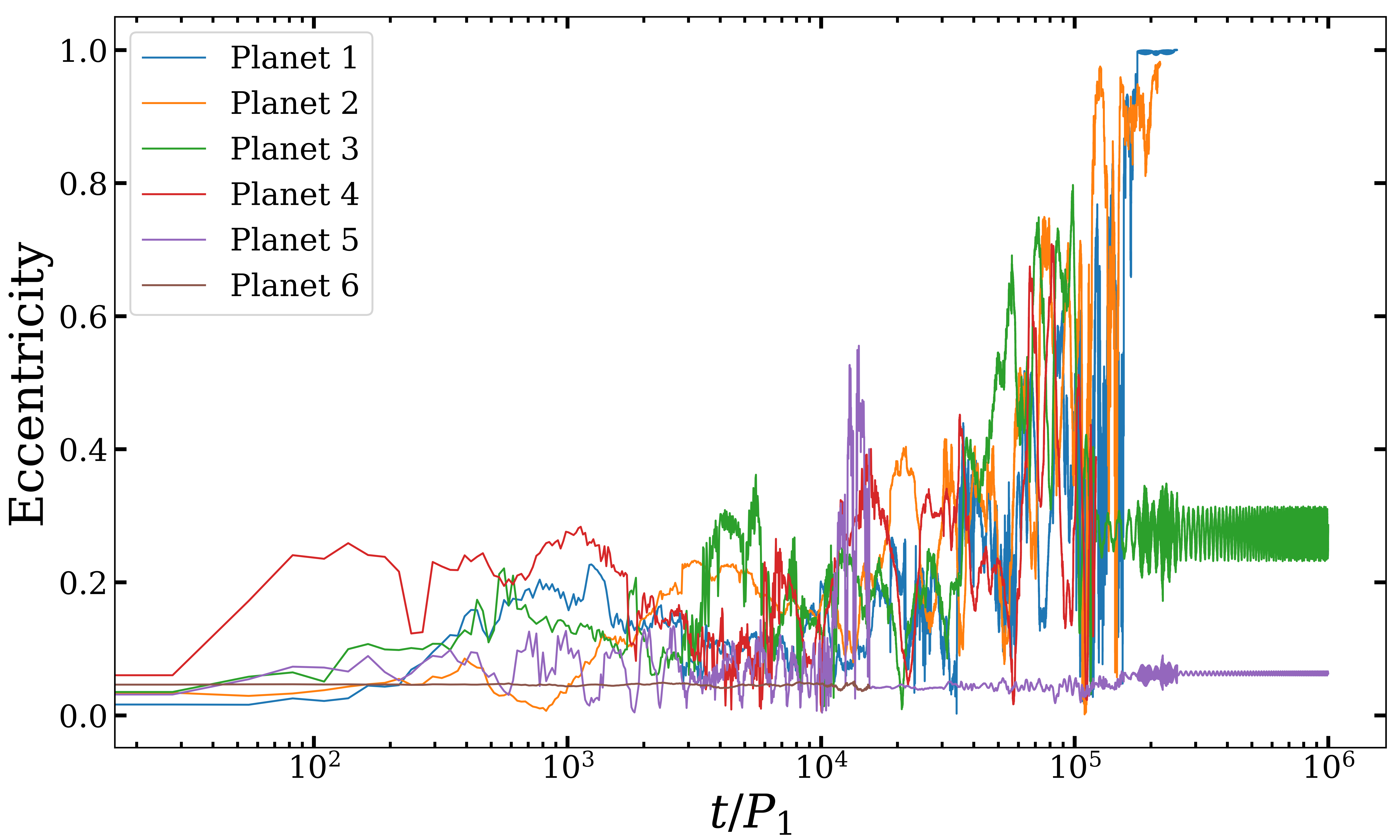}
    \includegraphics[width=0.45\linewidth]{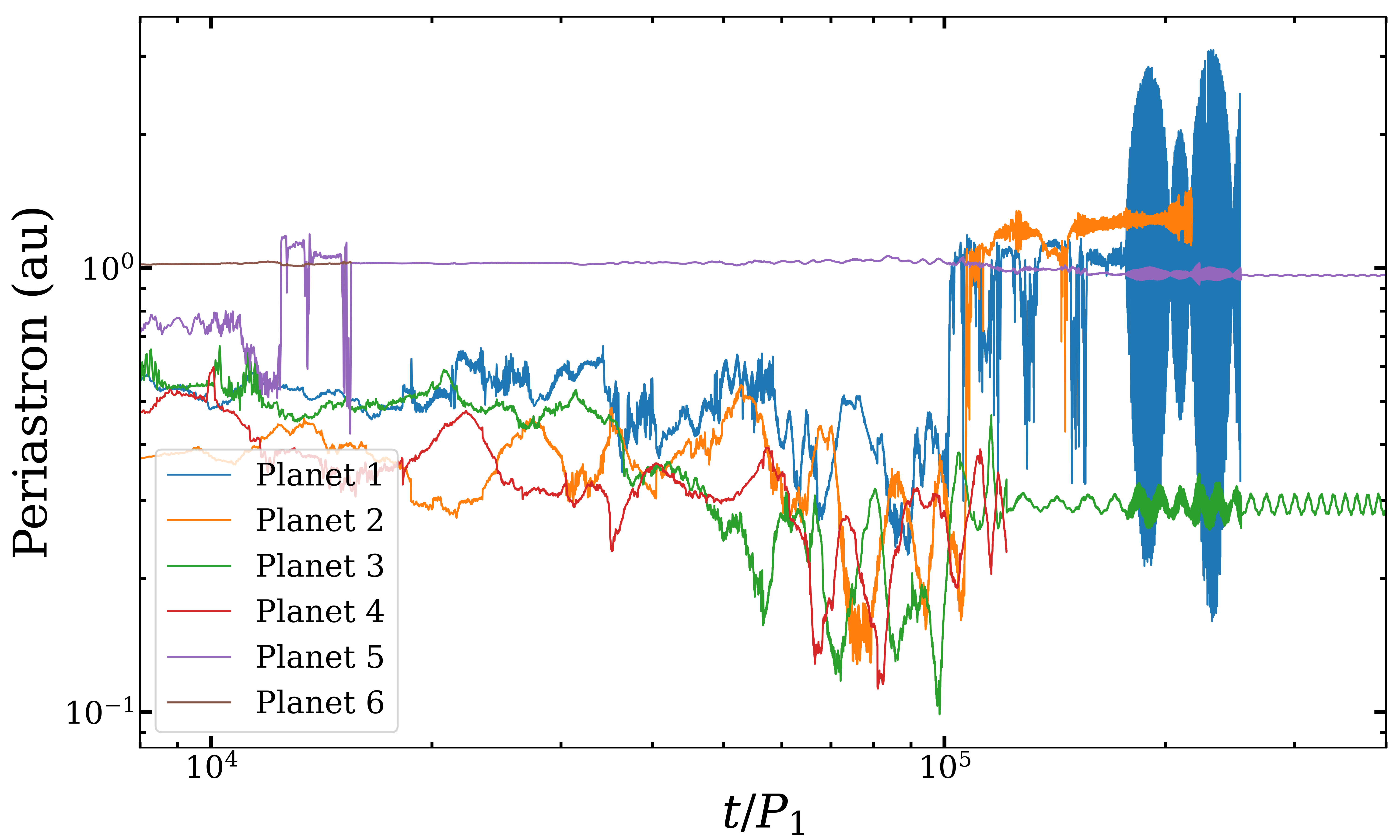}
    \caption{Evolution of the semimajor axes (upper left panel), eccentricities (lower left panel), apastron distances (upper right panel), and periastron distances (lower right panel) in a super-Earth-cold-Jupiter system. When a planet-planet collision happens, the new planet will be indexed by the planet with a smaller index after the merger. When a planet escapes, its curve will be terminated. At the start, the cold Jupiter is labeled Planet 6, while after the collision, it is labeled Planet 5. The left and right panels show different logarithmic ranges of $t/P_1$, so both early and late evolution can be seen clearly.}
    \label{fig:SEJ_evolution}
\end{figure*}

We choose a system that experiences 2 planet-planet collisions and 2 ejections as an example to show how a super-Earth-cold-Jupiter system evolves. Figure \ref{fig:SEJ_evolution} shows the evolution of the semimajor axis, eccentricity, periastron distance, and apastron distance of each planet. We index the planets from low to high in the order of their initial distances from the host, so Planet 6 is the cold Jupiter. Soon after the start of the integration, the orbits of the super-Earths become more eccentric. At $t/P_1 \sim 150$, Planets 1 and 2 have an encounter and switch order. Then at $t/P_1\sim 1.2\times 10^4$, Planet 5 is scattered into an orbit beyond that of the cold Jupiter, with its periastron near or inside the orbit of the cold Jupiter. After further changes in its orbit, it collides with the cold Jupiter at $t/P_1 \sim 1.5\times 10^4$, and the merged planet is labeled Planet 5. At $t/P_1 \sim 10^5$, Planets 1 and 2 are scattered into orbits farther from the host, with their periastrons near or inside the orbit of the cold Jupiter. Soon afterwards, Planet 3 collides and merges with Planet 4.  Eventually, Planets 1 and 2 eventually escape from the system, and Planet 3 and 5 form a stable super-Earth-cold-Jupiter system. This example shows the following typical events in the evolution of the super-Earths-cold-Jupiter systems: at the beginning, the orbits of the super-Earths become unstable gradually, but they cannot escape since their interactions with each other are relatively weak; when a super-Earth is scattered into an orbit that has strong interactions with the cold Jupiter; the cold Jupiter can collide with the super-Earth or transfer energy, resulting in the latter's ejection; the remaining super-Earths can collide with each other.

\begin{figure}[tb]
    \centering
    \includegraphics[width=1\linewidth]{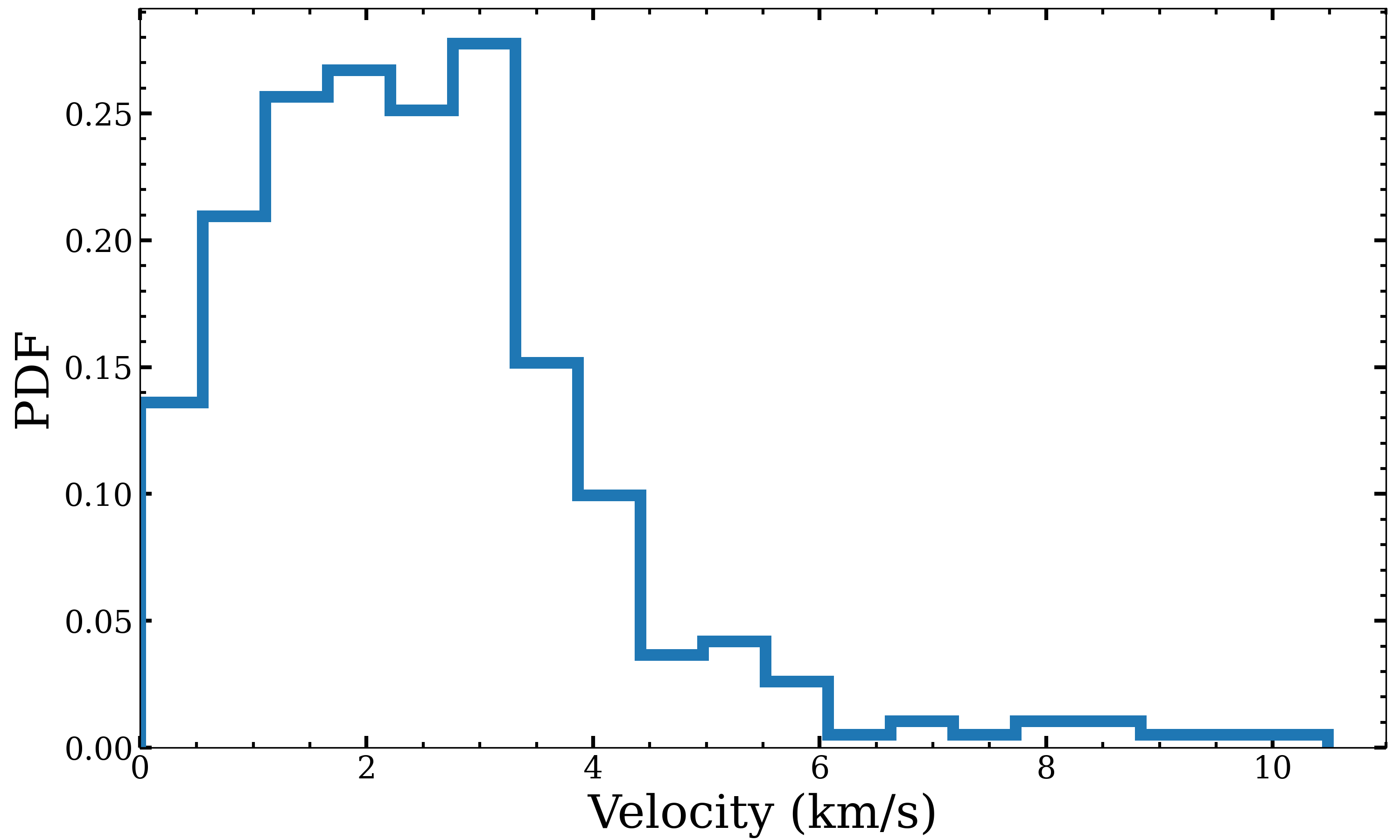}
    \caption{Distribution of the velocities of ejected planets relative to the host stars in the super-Earth-cold-Jupiter sample. Most ejected planets have relative velocities lower than 6 $\mathrm{km\ s^{-1}}$.}
    \label{fig:SEJ_velocity_dis}
\end{figure}

The velocity distribution of FFPs plays an essential role in their detection since it influences the timescales of their microlensing signals \citep{Paczynski1986}. Figure \ref{fig:SEJ_velocity_dis} shows a histogram of the velocities of ejected planets relative to the host stars in the super-Earth-cold-Jupiter sample. Most ejected planets have velocities in the range of $0\sim6\ \mathrm{km\ s^{-1}}$, implying that they obtain just enough kinetic energy to escape (the escape velocity at 1 au from a solar-mass star is about $40\ \mathrm{km\ s^{-1}}$). Given the high velocity dispersion of stars in the Milky Way ($\sim 120\ \mathrm{km\ s^{-1}}$ in the Galactic bulge \citep{Valenti2018} and $\sim 50\ \mathrm{km\ s^{-1}}$ in the Galactic disk \citep{Anguiano2020}), the combination of the velocity of the host stars and the relative velocity of the ejected planets would be dominated by the velocity of the stars. As a result, the velocity distribution of the ejected planets should be similar to that of stars.

\begin{figure*}[tb]
    \centering
    \includegraphics[width=0.4\linewidth]{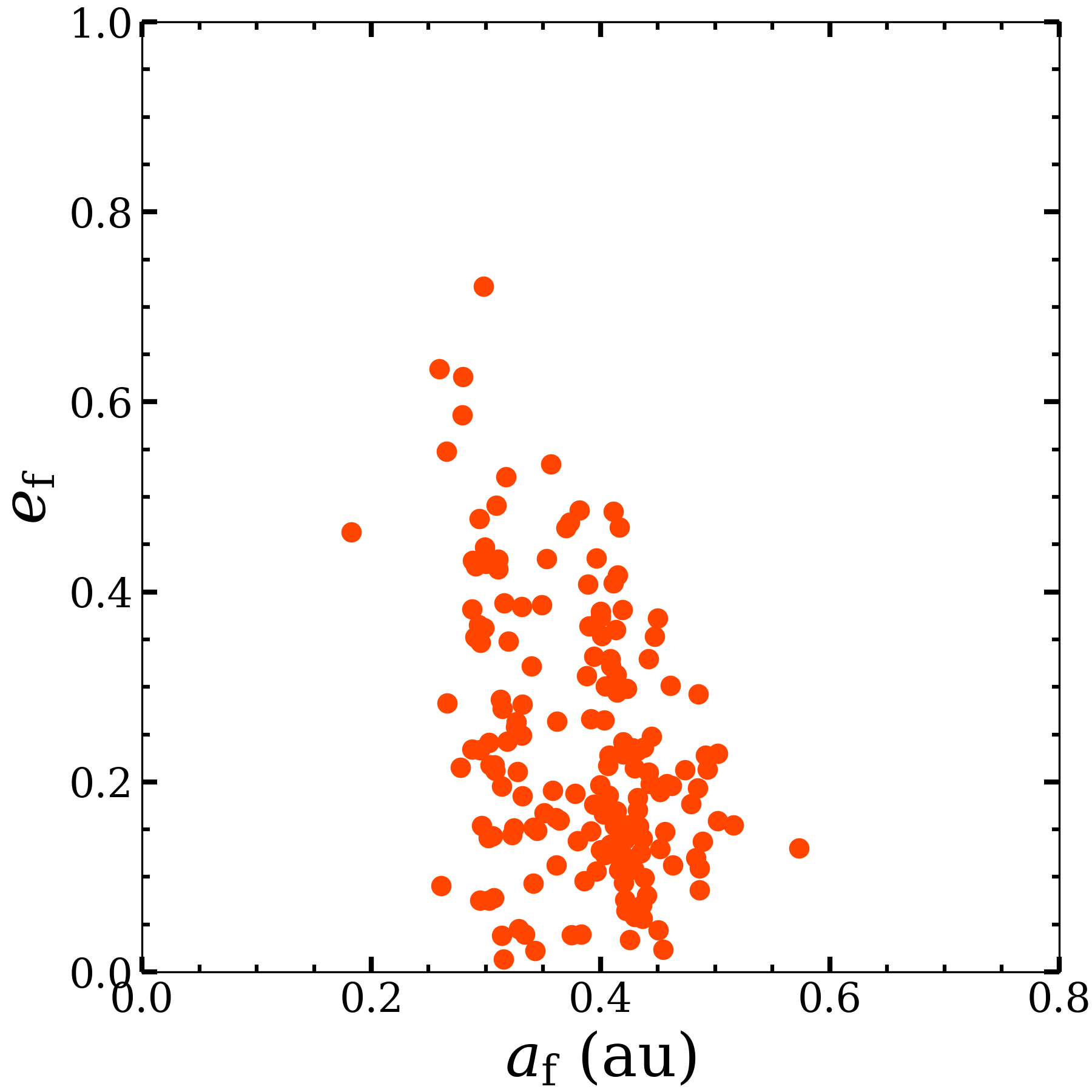}
    \includegraphics[width=0.4\linewidth]{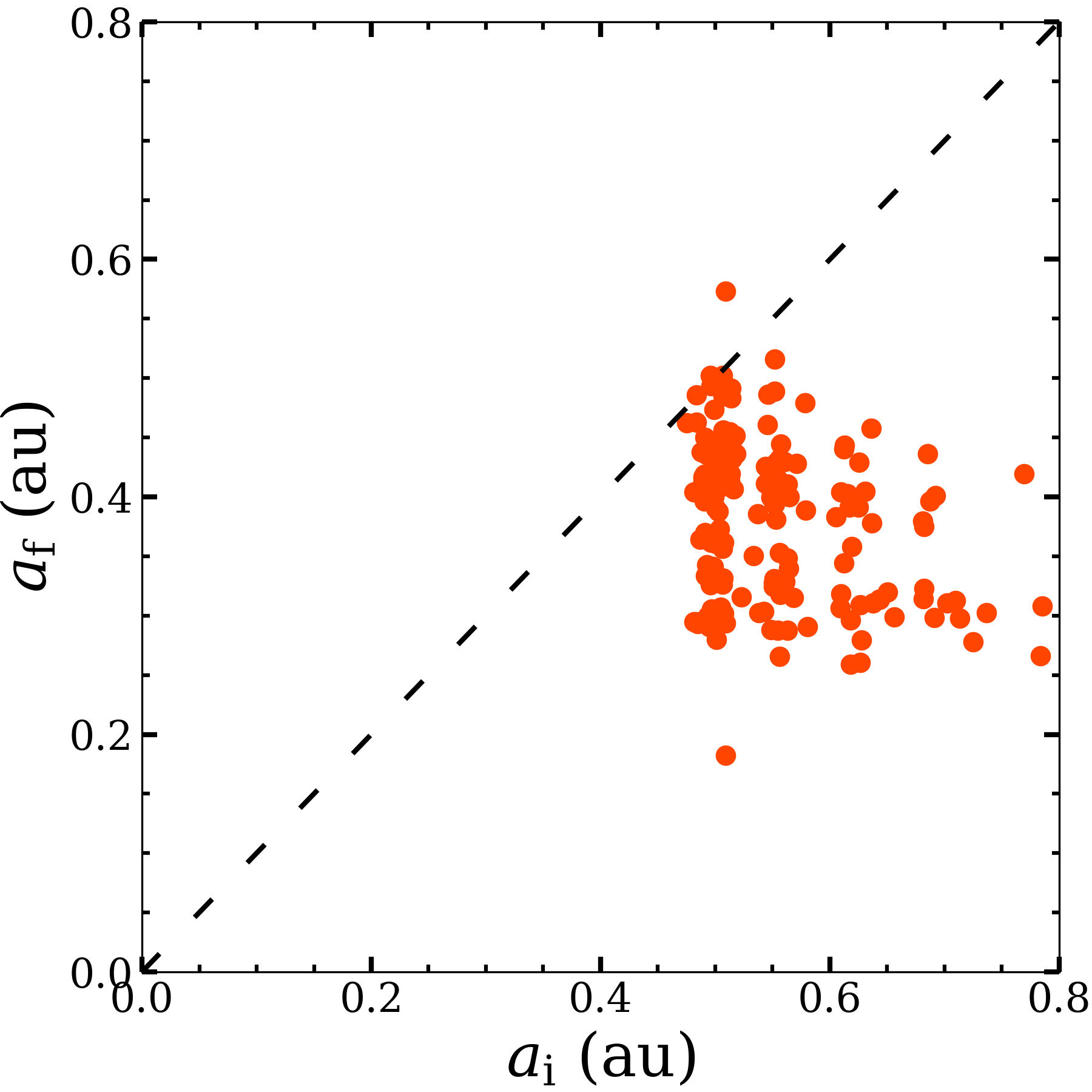}
    \includegraphics[width=0.4\linewidth]{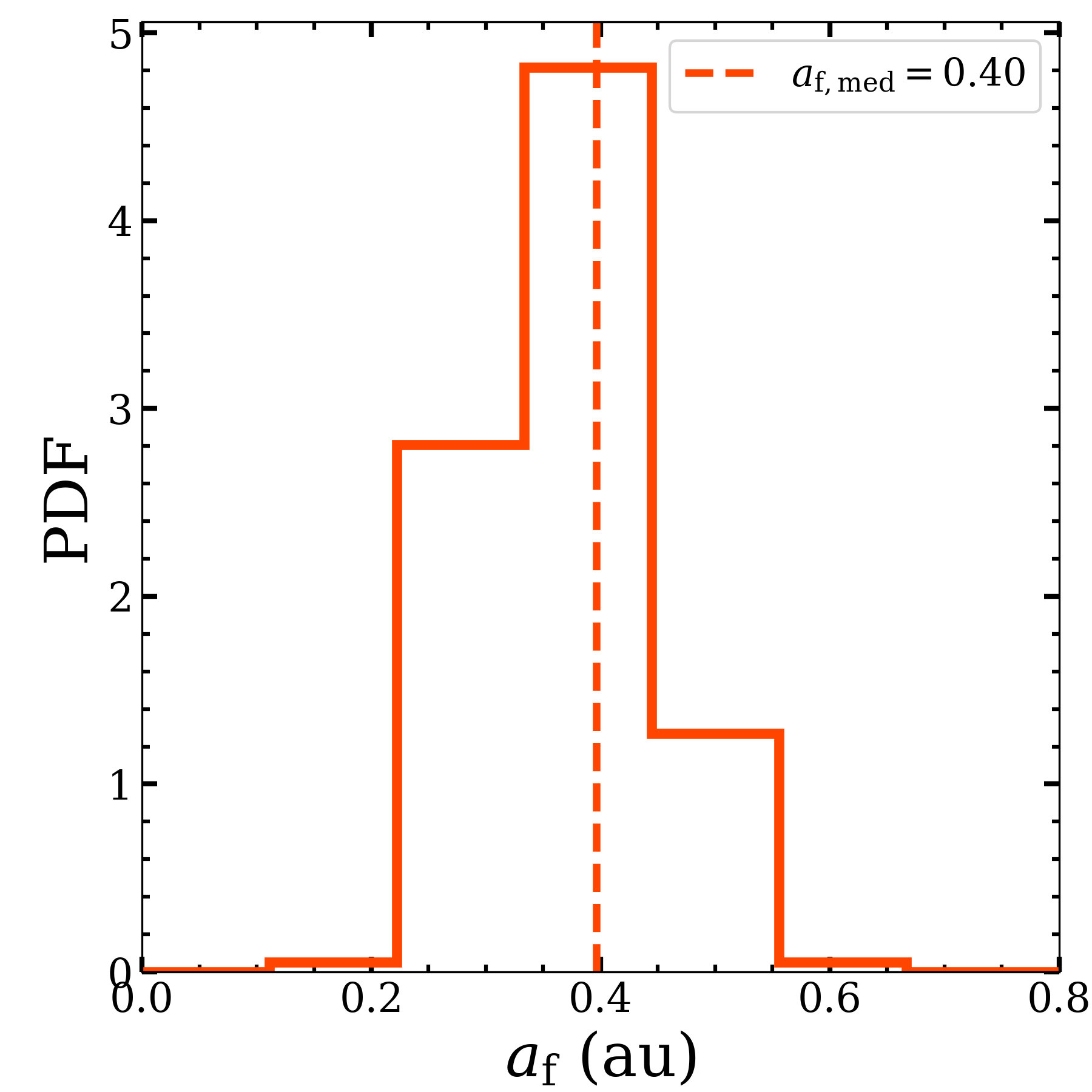}
    \includegraphics[width=0.4\linewidth]{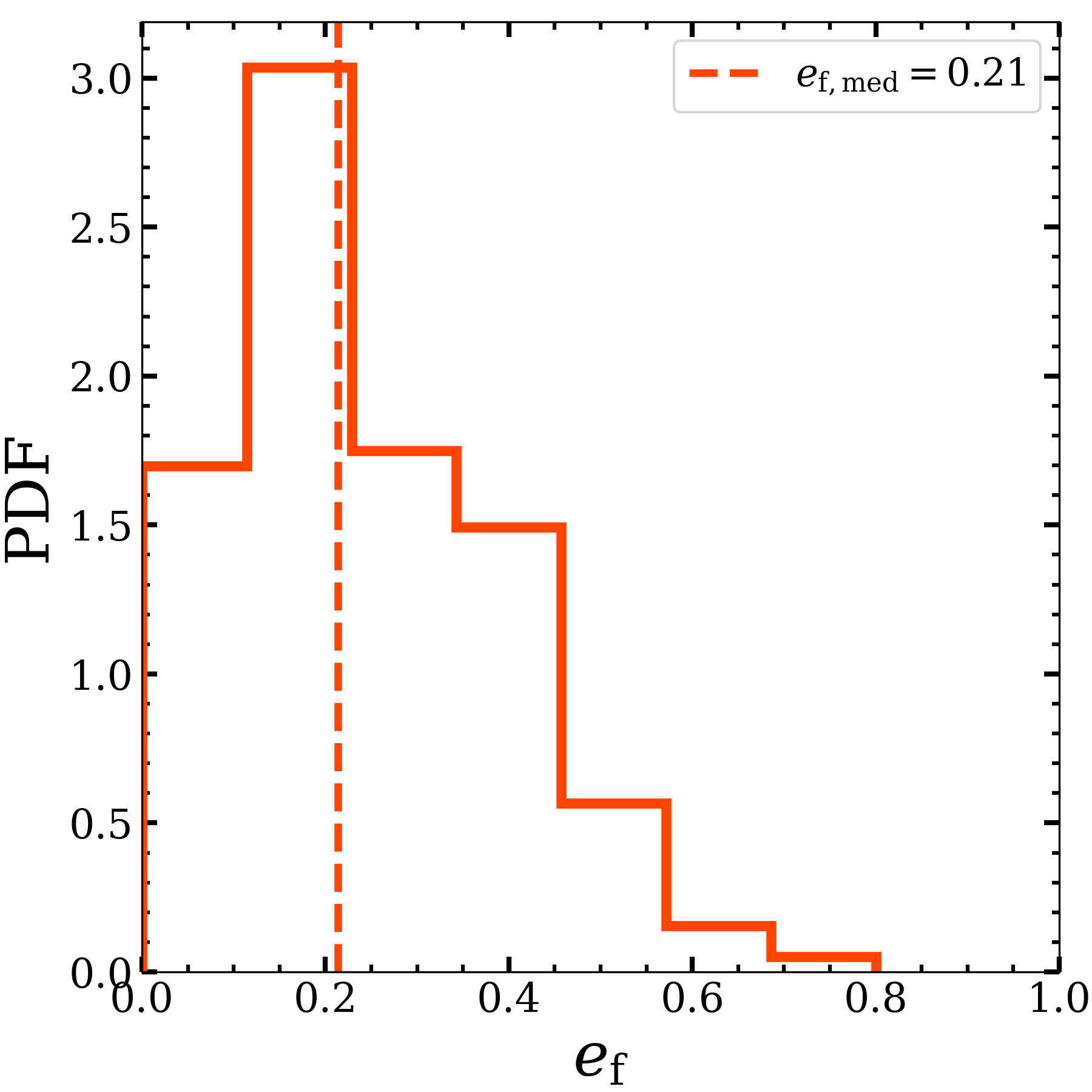}
    \caption{Orbital properties of the super-Earths in the final stable two-planet systems from the super-Earths-cold-Jupiter simulations. Top left panel: Final eccentricities vs.\ final semimajor axes. Top right panel: Final semimajor axes vs.\ initial semimajor axes. In this panel, the scatters are clustered in several strips of initial semimajor axes, corresponding to those of the 5 initial super-Earths. The super-Earths further away from the cold Jupiter are more likely to survive. Bottom left panel: Probability distribution of the final semimajor axes. Bottom right: Probability distribution of the final eccentricities.}
    \label{fig:SEJ_af_scatter_hist}
\end{figure*}

At the end of the simulations, 94\% of the stable planetary systems are two-planet systems composed of a super-Earth and a cold Jupiter. Since the orbit of the cold Jupiter is only weakly perturbed by the super-Earths, we study the properties of the super-Earth orbit in the stable two-planet systems.
The top panels of Figure \ref{fig:SEJ_af_scatter_hist} show the final eccentricities vs.\ the final semimajor axes and the final semimajor axes vs.\ the initial semimajor axes. The orbits of many remaining super-Earths have significant eccentricities: the smaller the final semimajor axes, the greater the range of eccentricities. In addition, most remaining super-Earths have semimajor axes smaller than 0.5 au, meaning their final semimajor axes are smaller than the initial ones. In the bottom panels of Figure \ref{fig:SEJ_af_scatter_hist}, we show the probability distribution of the final semimajor axes and final eccentricities of the surviving super-Earths. The final semimajor axes are mostly in the $0.2\sim0.6\ \mathrm{au}$ range, with a median of $a_\mathrm{f,med} = 0.40\ \mathrm{au}$, which is smaller than the initial semimajor axes of the super-Earths. At the same time, the orbits have become more eccentric. The median of the final eccentricities is $e_\mathrm{f,ave} = 0.21$, and the 90th percentile value is $e_\mathrm{f,90} = 0.44$ (see Table \ref{tab:SEJ_K_stats}), which are much greater than the initial $[0.01,0.05]$ range.
These results indicate that in the super-Earths-cold-Jupiter systems, the remaining super-Earths move inwards to survive since, in this way, they are less likely to interact with the cold Jupiters, which decreases the probability of collision or ejection. 

A general picture of the dynamical evolution of the super-Earth-cold-Jupiter systems is as follows: orbits of the super-Earths become chaotic under the interactions with each other and with the cold Jupiter. Some super-Earths move outwards and interact with the cold Jupiter. In the meantime, others move inwards. Eventually, the super-Earths moving outwards are lost through collision or ejection after close encounters with the cold Jupiter, while the super-Earths moving inwards remain because of their considerably larger distance from the cold Jupiter. The orbits of these surviving super-Earths become more eccentric from the frequent interactions.

\section{Discussion}\label{sec:discussion}
\subsection{Orbital Spacing $K$}\label{sec:Spacing}

Previous studies on the dynamical instability of planetary systems discovered that the initial orbital spacing $K$, defined by Eq.~(\ref{eq:separation}), influences the first close encounter time. The larger the value of $K$, the later the first encounter happens \citep[e.g.,][]{Chambers1996, Smith2009, Rice2023}. On the other hand, in Section \ref{sec:superearth}, we find from the super-Earths only simulations with $K = 5$ that the first close encounter time $\tilde{t}_\mathrm{FCE}$ and the half-loss time $\tilde{t}_{1/2}$ (both in units of $P_1$) have different dependence on the initial semimajor axis of the innermost orbit $a_{1,0}$, with $\tilde{t}_\mathrm{FCE}$ nearly independent of $a_{1,0}$ and $\tilde{t}_{1/2} \sim a_{1,0}^b$, where $b \approx 2.1$. Thus, we study further the influence of the orbital spacing $K$ on $\tilde{t}_\mathrm{FCE}$ and $\tilde{t}_{1/2}$ by extending our simulations of the super-Earths only systems to $K = 4$ and 6.

\begin{figure}[bt]
    \centering
    \includegraphics[width = 1\linewidth]{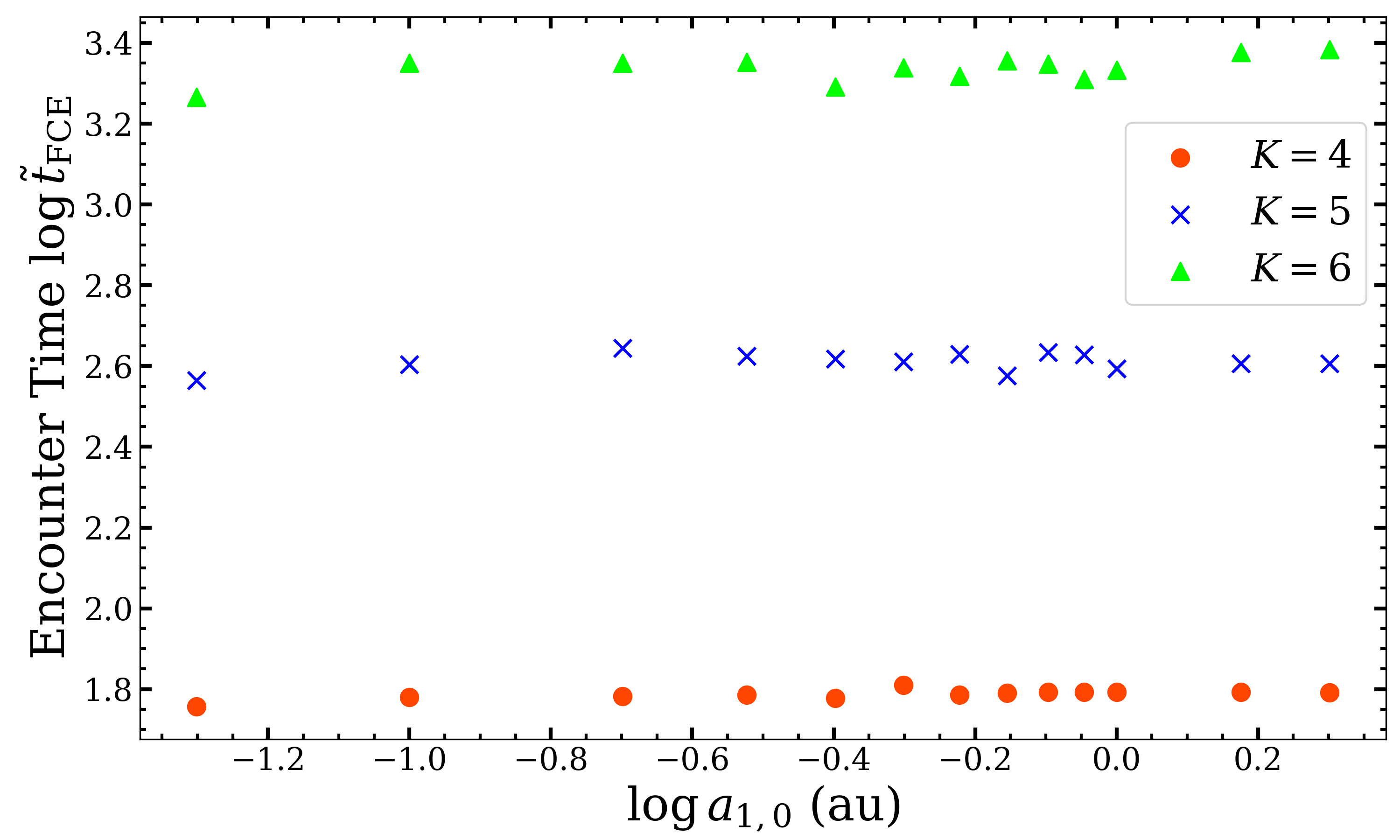}
    \caption{The first close encounter time $\tilde{t}_\mathrm{FCE}$ vs.\ the initial semimajor axis of the innermost orbit $a_{1,0}$ for super-Earths only simulations with orbital spacing $K = 4$, 5, and 6, defined in Equation (\ref{eq:separation}).}
    \label{fig:superearth_encounter_t_K}
\end{figure}

\begin{figure}[bt]
    \centering
    \includegraphics[width = 1\linewidth]{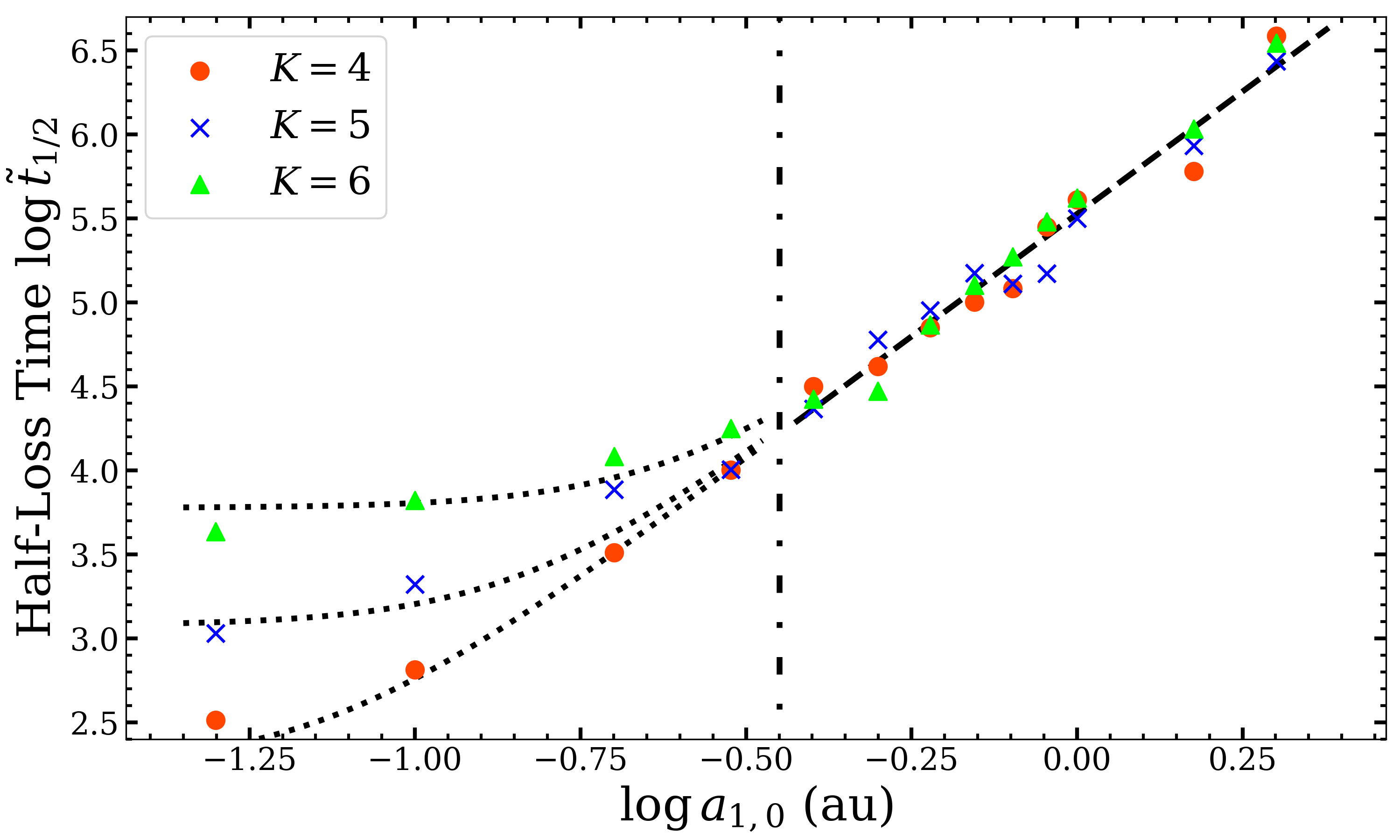}
    \caption{The half-loss time $\tilde{t}_{1/2}$ vs.\ the initial semimajor axis of the innermost orbit $a_{1,0}$ for super-Earths only simulations with orbital spacing $K = 4$, 5, and 6. The vertical dash-dotted line is $\log a_{1,0} = -0.45$, which roughly divides the dependence of $\tilde{t}_{1/2}$ on $K$ into two parts. The dashed and dotted lines show Equations (\ref{eq:half_loss_t}) and (\ref{eq:half_loss_t_2}), respectively.}
    \label{fig:superearth_half_t_K}
\end{figure}

In Figures \ref{fig:superearth_encounter_t_K} and \ref{fig:superearth_half_t_K}, we show the first close encounter time $\tilde{t}_\mathrm{FCE}$ and the half-loss time $\tilde{t}_{1/2}$, respectively, for different $\log a_{1,0}$ and $K$. The first close encounter time $\tilde{t}_\mathrm{FCE}$ is nearly independent of $a_{1,0}$ for a given $K$ (as we find in Section \ref{sec:superearth} for $K = 5$) and increases rapidly with increasing $K$ (as found in previous studies).
For all values of $K$, the half-loss time $\tilde{t}_{1/2}$ increases with $a_{1,0}$. The vertical dash-dotted line in Figure \ref{fig:superearth_half_t_K} is $\log a_{1,0} = -0.45$, roughly dividing the figure into two parts. For $\log a_{1,0} > -0.45$, $\tilde{t}_{1/2}$ of different $K$ at the same $a_{1,0}$ does not have significant difference. When we linearly regress $\log \tilde{t}_{1/2}$ to $\log a_{1,0}$ of all data points whose $\log a_{1,0} > -0.45$, we obtain:
\begin{equation}\label{eq:half_loss_t}
    \log \tilde{t}_{1/2} = (2.92 \pm 0.11)\log a_{1,0} + (5.53 \pm 0.02),
\end{equation}
(see dashed line in Figure \ref{fig:superearth_half_t_K}). For $\log a_{1,0} < -0.45$, $\tilde{t}_{1/2}$ is larger than the extrapolation of Equation (\ref{eq:half_loss_t}), with the deviation larger at smaller $a_{1,0}$ and larger $K$. This can be understood when we realize that the first planet lost cannot happen before the first close encounter, i.e., $\tilde{t}_{1/2}$ must be larger than $\tilde{t}_\mathrm{FCE}$, and that $\tilde{t}_\mathrm{FCE}$ increases rapidly with $K$ (Figure \ref{fig:superearth_encounter_t_K}). In fact, the data points in Figure \ref{fig:superearth_half_t_K} can be approximated by
\begin{equation}\label{eq:half_loss_t_2}
\tilde{t}_{1/2} = 10^{5.53} a_{1,0}^{2.92} + \ell \tilde{t}_\mathrm{FCE} , 
\end{equation}
where $\ell \approx 3$ (see dotted lines in Figure \ref{fig:superearth_half_t_K} for $K=4$, 5, and 6).
 
We also vary the orbital spacing of the super-Earth-cold-Jupiter samples from $K = 5$ to $K =4$ and 6, with fixed $a_{1,0} = 0.5$ au. We find that the value of $K$ does not change the statistical properties of the results (see Table \ref{tab:SEJ_K_stats}). The reason may be that the outcomes (ejection, collision, and scattering) of the encounters between the super-Earths and the cold Jupiter depend primarily on the Safronov number (see next Section), which is not sensitive to the value of $K$.

\subsection{Safronov Number and Ejection}\label{sec:Safronov}

In the super-Earths only sample, the number of ejections increases with the distance between the planets and the host. The $a_{1,0}=0.1\ \mathrm{au}$ sample has no ejection, but the $a_{1,0}=2\ \mathrm{au}$ sample has 15\% of super-Earths ejected (Figure \ref{fig:superearth_ejection_fraction}). The probability that a planet is going to escape should depend on the Safronov number \citep{Safronov1972}:
\begin{equation}\label{eq:safronov_number}
    \Theta = \frac{1}{2}\frac{V^2_\mathrm{esc}}{V^2_\mathrm{orb}} = \left(\frac{M_\mathrm{p}}{M_\star}\right) \left(\frac{a_\mathrm{p}}{r_\mathrm{p}}\right),
\end{equation}
where $V_\mathrm{esc}$ is the escape velocity at the planet surface, $V_\mathrm{orb}$ is the orbital velocity of the planet, $a_\mathrm{p}$ is the semimajor axis of the planet, and $r_\mathrm{p}$ is the planet radius. When two planets get close to each other, they are likely to gain velocity changes at the order of $V_\mathrm{esc}$. Hence, a greater $V_\mathrm{esc}$ adding to $V_\mathrm{orb}$ makes planets easier to escape \citep{Morbidelli2018}. When $\Theta < 1$ ($\Theta > 1$), planets tend to collide (escape). 

To study the relationship between the Safronov number and ejection, we simulate systems with equal-mass giant planets with greater Safronov numbers than super-Earths with the same innermost semimajor axis. The planet masses are Jupiter, Saturn, and $1/3$ Saturn mass. The planet radii are Jupiter radius, Saturn radius, and $(1/3)^{1/3}$ Saturn radius. Other initial conditions of the numerical simulations are shown in Table \ref{tab:equal_mass_initial}.

\begin{table}
    \centering
    \caption{Initial Conditions of Equal-mass Planet Simulations}
    \label{tab:equal_mass_initial}
    \begin{tabular}{l c r}
        \hline
        \hline
        Number of planets & \quad & 5 equal-mass planets \\
        Planet mass $m_\mathrm{p}$ & \quad & $M_\mathrm{J}$, $M_\mathrm{S}$, or 1/3 $M_\mathrm{S}$ \\
        Planet radius $r_\mathrm{p}$ & \quad & $R_\mathrm{J}$, $R_\mathrm{S}$, or $(1/3)^{1/3} R_\mathrm{S}$ \\
        $a_1$ & \quad & Gaussian distribution of $a_{1,0} \pm 0.01\ \mathrm{au}$ \\
        $K$ & \quad & 5 \\
        \hline
    \end{tabular}
\tablecomments{$M_\mathrm{J}$ and $M_\mathrm{S}$ are the masses of Jupiter and Saturn, respectively, and $R_\mathrm{J}$ and $R_\mathrm{S}$ are the corresponding radii.}
\end{table}

\begin{figure}[bt]
    \centering
    \includegraphics[width = 1\linewidth]{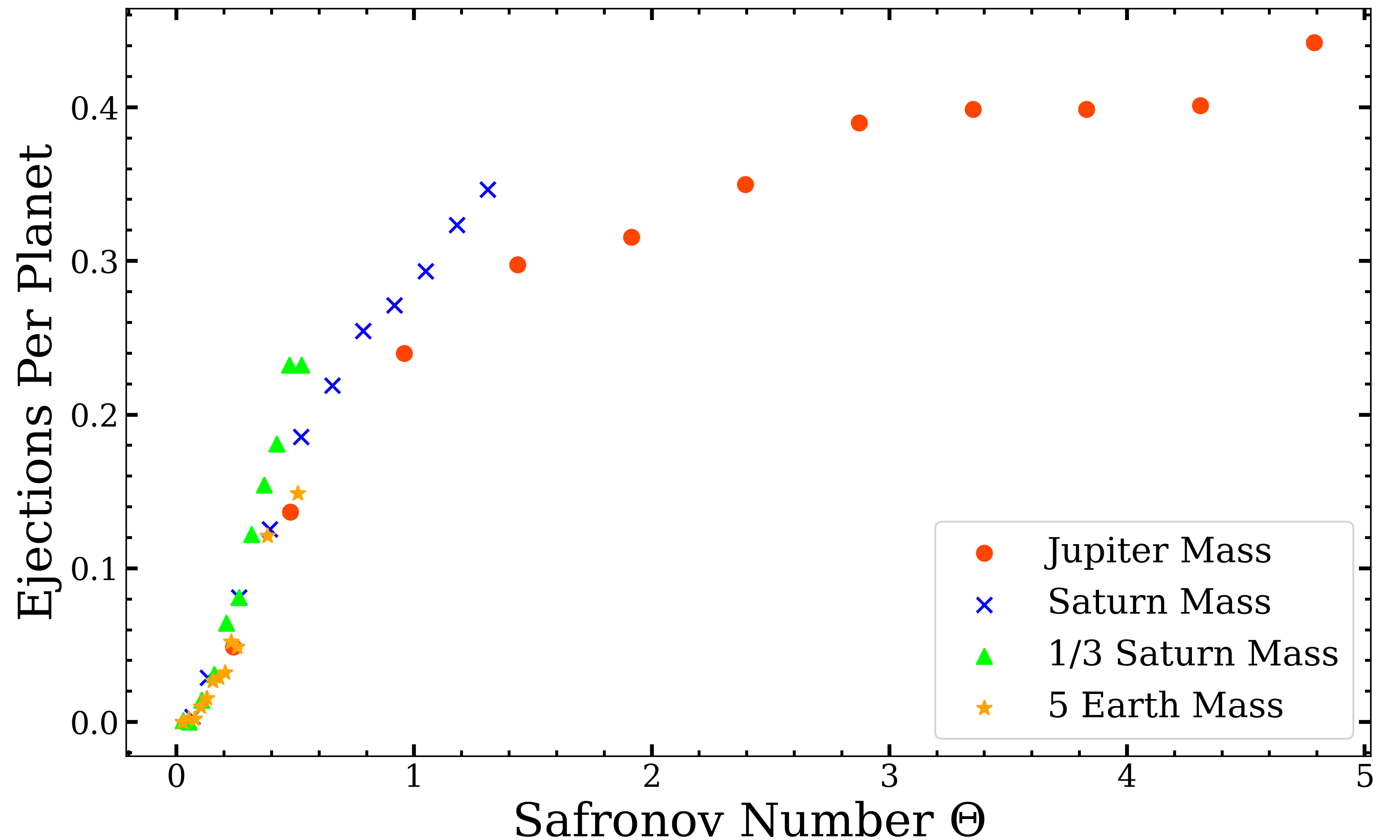}
    \caption{The number of ejections per planet vs.\ Safronov number, defined in Eq. (\ref{eq:safronov_number}), in the equal-mass planet samples together with the super-Earths only sample. The Safronov number of a sample is calculated based on the orbit of the middle (third) planet.}
    \label{fig:ejections_a_multi}
\end{figure}

\begin{figure}[bt]
    \centering
    \includegraphics[width = 1\linewidth]{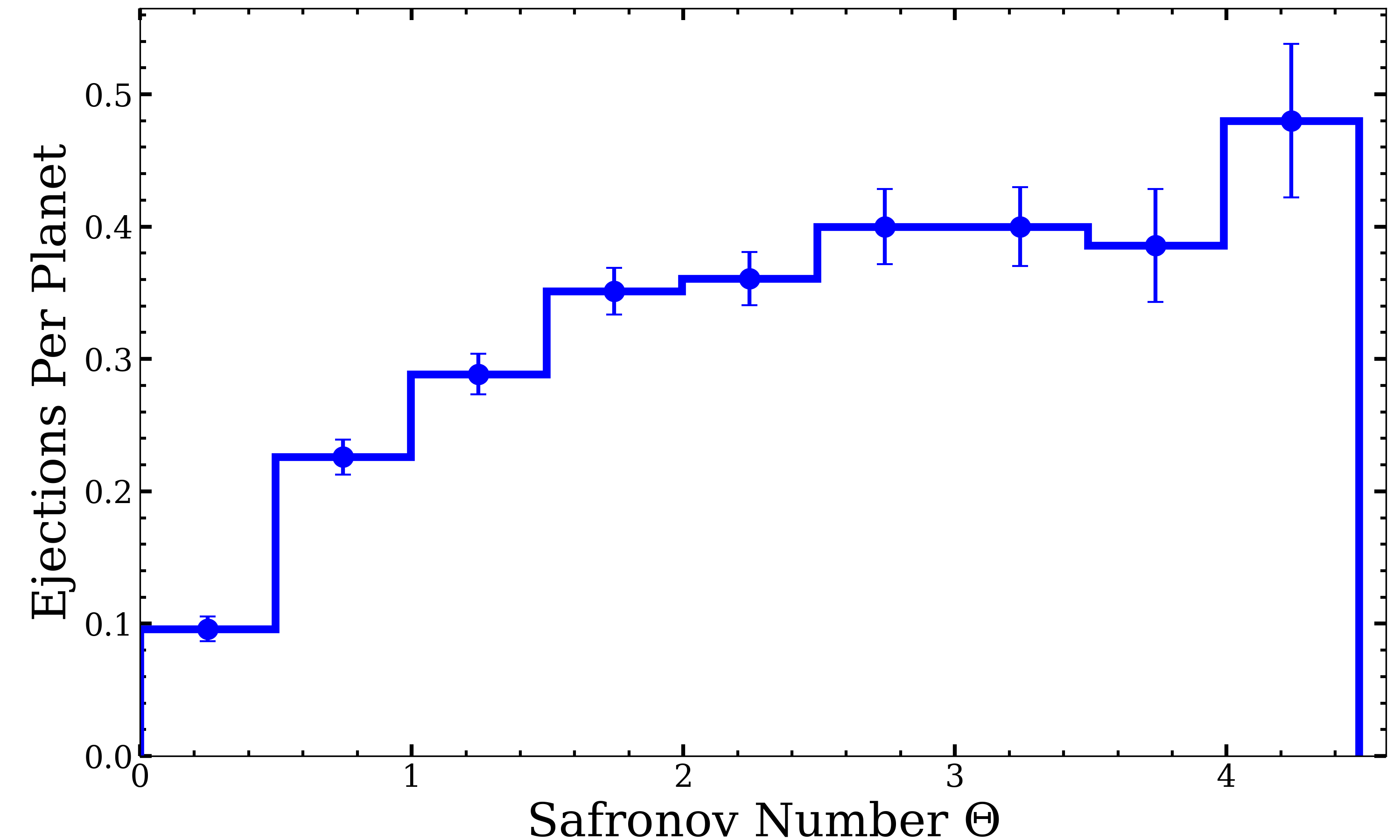}
    \caption{Histogram of the number of ejections per planet vs.\ Safronov number in the equal-mass planet sample with randomly chosen masses and semimajor axis of the innermost orbit.}
    \label{fig:ejections_safronov}
\end{figure}

In Figure \ref{fig:ejections_a_multi}, we show the number of ejections per planet vs.\ Safronov number in the equal-mass planet and super-Earth samples. In each sample, the Safronov number is defined as that of the middle (third) planet. We see that samples of different masses but similar Safronov numbers agree well on the probability of ejection, which implies that the Safronov number is a good indicator of the tendency of a planetary system to undergo collision or ejection. On this basis, we further simulate equal-mass planetary systems with the planet mass randomly chosen in $[1/30, 1]M_\mathrm{J}$ and the semimajor axis of the innermost orbit in $[0.1,1]$\,au. We set the radii of the planets by keeping their density the same as Jupiter. Initial conditions are given in Table \ref{tab:safronov_initial}. Figure \ref{fig:ejections_safronov} shows that the probability of planet ejection increases with the Safronov number, especially when the Safronov number is small.

\begin{table}
    \centering
    \caption{Initial Conditions of Equal-mass Planet Simulations with Randomly Chosen Masses and Semimajor Axis of the Innermost Orbit}
    \label{tab:safronov_initial}
    \begin{tabular}{l c r}
        \hline
        \hline
        Number of planets & \quad & 5 equal-mass planets \\
        Planet mass $m_\mathrm{p}$ & \quad & Randomly chosen in $[1/30, 1]M_\mathrm{J}$ \\
        Planet radius $r_\mathrm{p}$ & \quad & $(m_p/M_\mathrm{J})^{1/3}R_\mathrm{J}$ \\
        $a_1$ & \quad & Randomly chosen in $[0.1,1]$ au \\
        $K$ & \quad & 5 \\
        \hline
    \end{tabular}
\end{table}

As we mentioned in Section \ref{sec:intro}, \citet{Matsumoto2017}, \citet{Rice2018}, and \citet{Bartram2021} have performed simulations that extend to the first planet loss event and beyond, but they did not find any planet ejection events in their simulations. The absence of planet ejection
in the simulations by \citet{Matsumoto2017} and \citet{Bartram2021} is consistent with the fact that the Safronov numbers of their simulations are much less than one. For the simulations with four planets of mass $m = 1 \times 10^{-5} M_\sun$ and bulk density of $2\,$g$\,$cm$^{-3}$ by \citet{Rice2018}, the Safronov number $\Theta \approx 0.11 (a_{1,0}/\text{au})$, which would be $\approx 1.1$ and $11$ for the simulations with $a_{1,0} = 10$ and $100\,$au, respectively. Thus the absence of planet ejection or collision with the central star in these simulations, as reported by \citet{Rice2018}, is unexpected on the basis of their Safronov number, and additional analysis of simulations with similar parameters will be needed.

A study of FFPs in the Galactic bulge conducted by Microlensing Observations in Astrophysics (MOA) found a large number of FFPs: if the planet mass distribution is assumed to be a power-law, then the number of FFPs per star is $f = 21^{+23}_{-13}$ with a total mass of $m = 80^{+73}_{-47} M_\oplus$ \citep{Sumi2023}. The uncertainty of this result is still large due to the small sample size. If further studies confirm this high abundance, one possibility may be that many planets are formed in protoplanetary disks and become unstable to become FFPs. In our simulations, we find about 40\% probability that a cold Jupiter can eject a super-Earth, and also 40\% probability that equal-mass planets can escape when they are far away from their host stars. Many planets of small masses might be ejected right after planetary systems form. However, a more detailed prediction will require more realistic initial conditions of the planetary systems, starting with many planetary embryos. This is beyond the scope of the current paper.

\subsection{Comparison with Observations}

\begin{figure}[bt]
    \centering
    \includegraphics[width = 1\linewidth]{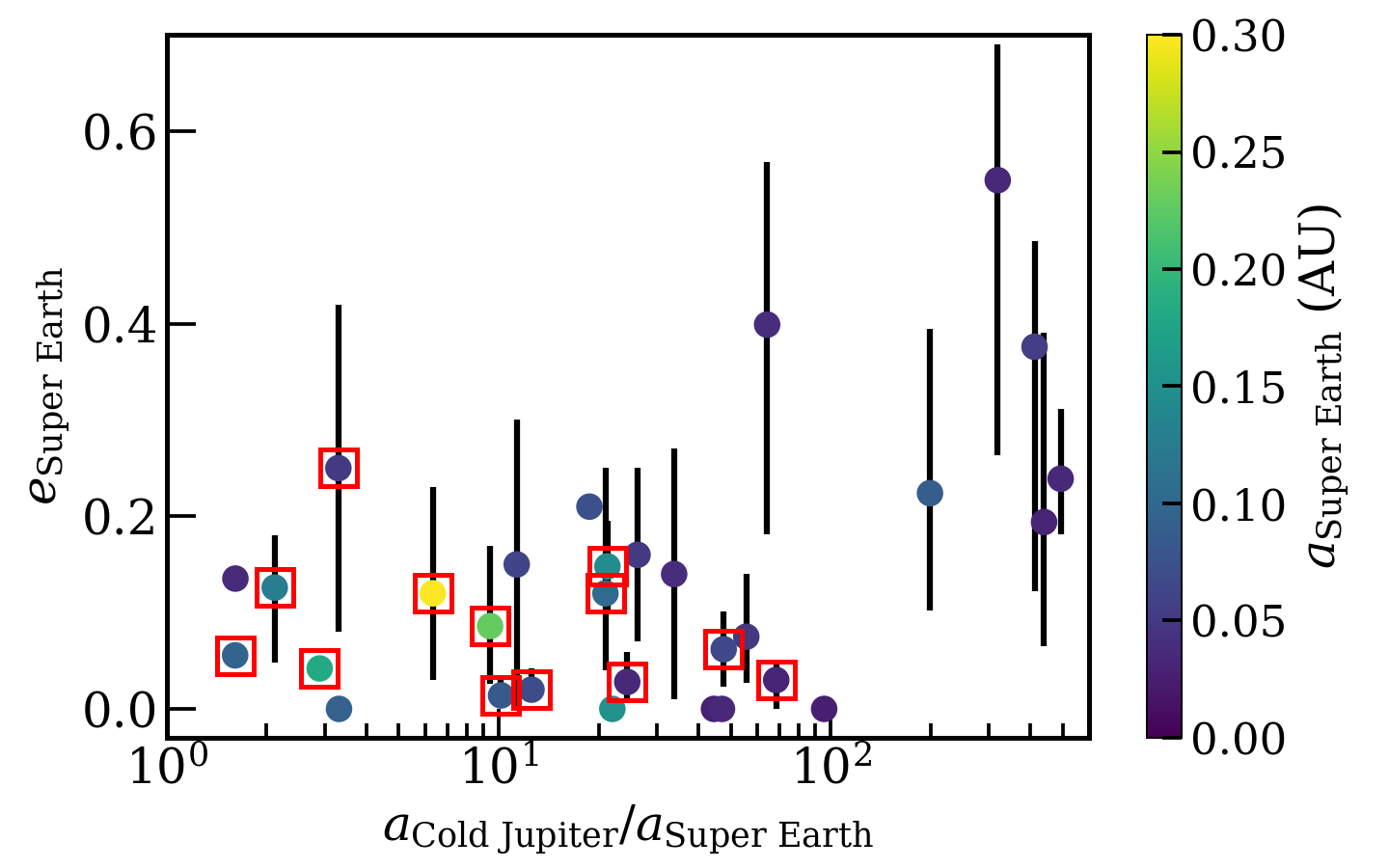}
    \caption{Eccentricity of inner super-Earth vs.\ the semimajor axis ratio between the outer cold Jupiter and the inner super-Earth from observations, with the color of the symbol showing the semimajor axis of the inner super-Earth. The red boxes mark the systems with multiple inner super-Earths. }
    \label{fig:observations}
\end{figure}

For comparison with observed systems, we retrieve a list of confirmed planetary systems with a single outer gas giant accompanied by inner super-Earths from the NASA Exoplanet Archive\footnote{\url{https://exoplanetarchive.ipac.caltech.edu}} \citep{Akeson2013}. We define gas giants as planets with mass $m \geq 0.3\ {\rm M_{J}}$ while restricting super-Earths to planets with mass below $15\ {\rm M_{\oplus}}$. We select systems as long as the giant planet is outside the orbit of the super-Earth, regardless of its orbital period, and we only accept those that have eccentricity measurements on the inner super-Earths. In the end, our observational sample consists of 16 systems containing a single inner super-Earth, with the semimajor axes of the outer giant planets between 0.3 and 22 au, most of which are at 1--4 au. In addition, we also identify 13 inner super-Earths (with eccentricity measurements) belonging to 8 systems with multiple super-Earths plus an outer giant planet, which have properties similar to the single-super-Earth systems above. Figure~\ref{fig:observations} presents the eccentricity of the inner super-Earths as a function of the semimajor axis ratio between the inner super-Earth and the outer giant planet. We find that the majority of systems have inner super-Earths located at 0.1--0.5 au and large ratio of the semimajor axis of the cold Jupiter to the super-Earth(s).

Based on the properties of these observed systems, we simulate systems with smaller innermost semimajor axis $a_{1,0}$ of super-Earths compared to the super-Earth-cold-Jupiter systems in Section \ref{sec:SECJ}, while keeping the cold Jupiter at 1 au. The orbital spacing of the super-Earths is $K=5$.

Table \ref{tab:f_ej_se_a} shows that when the super-Earths start in the inner regions of planetary systems, they are less likely to be ejected and the fraction of systems with only two planets remaining decreases. The cold Jupiter cannot interact with the super-Earths effectively over this long distance. The evolution resembles the super-Earths only systems. In contrast, the super-Earths with initial orbits closer to the cold Jupiter experience significantly more ejections. It indicates that free-floating super-Earths and super-Earths coexisting with cold Jupiters may originate from different regions in planetary systems.


\begin{table}
    \centering
    \caption{Statistical Outcomes of Super-Earth-Cold-Jupiter Simulations with Cold Jupiter at 1 au, $K = 5$ for the super-Earths, and Different $a_{1,0}$}\label{tab:f_ej_se_a}
    \begin{tabular}{l | c c c c c}
        \hline
        \hline
        $a_{1,0}$ (au) & 0.1 & 0.2 & 0.3 & 0.4 & 0.5 \\
        $f_\mathrm{2p}$ & 0\% & 16\% & 66\% & 92\% & 96\% \\
        $f_\mathrm{ej, SE}$ & 0\% & 3\% & 20\% & 32\% & 37\% \\
        \hline
    \end{tabular}
\tablecomments{$f_\mathrm{2p}$ is the fraction of systems with two planets remaining. $f_\mathrm{ej, SE}$ is the fraction of all super-Earths that have been ejected.}
\end{table}

\subsection{Stability of Surviving Two-planet Systems}

In the super-Earth-cold-Jupiter sample in Section \ref{sec:SECJ} with $a_{1,0} = 0.5\,$au and $K = 5$ for all planets, more than 90\% of the systems end with 2 planets, a super-Earth and a cold Jupiter. In Figure \ref{fig:SEJ_frac}, at the end of Phase 2, the fractions of systems with different numbers of planets have become stable, an indicator that the systems are stable. We can use the empirical criterion \citep{Petrovich2015} for two-planet systems to determine their stability. The stability boundary is:
\begin{equation}\label{eq:sta_boundary}
    r_\mathrm{ap} \equiv \frac{a_\mathrm{out}(1 - e_\mathrm{out})}{a_\mathrm{in}(1 + e_\mathrm{in})} > Y,
\end{equation}
where $a_\mathrm{in}$ and $a_\mathrm{out}$ are the orbital semimajor axes of the inner and outer planets, respectively, $e_\mathrm{in}$ and $e_\mathrm{out}$ are the corresponding orbital eccentricities, and \begin{equation}\label{eq:Y}
    Y = 2.4\left[\frac{\mathrm{max}(m_\mathrm{in},\ m_\mathrm{out})}{M_\star}\right]^{1/3}\left(\frac{a_\mathrm{out}}{a_\mathrm{in}}\right)^{1/2} + 1.15 .
\end{equation}
In Equation (\ref{eq:Y}), $m_\mathrm{in}$ and $m_\mathrm{out}$ are the masses of the inner and outer planets, respectively, and $M_\star$ is the mass of the host star.

\begin{figure}[tb]
    \centering
    \includegraphics[width = 1\linewidth]{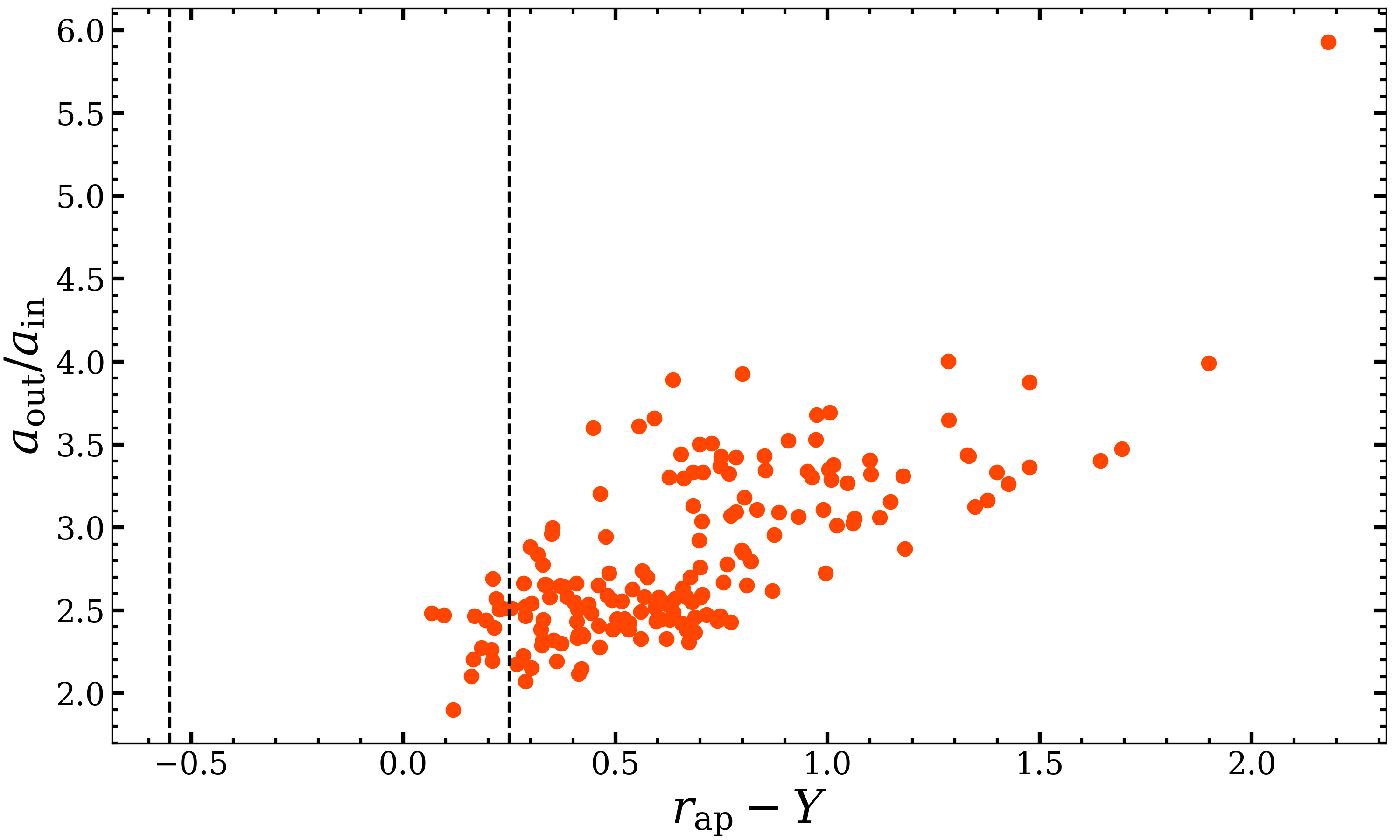}
    \caption{$a_\mathrm{out} / a_\mathrm{in}$ vs.\ $r_\mathrm{ap} - Y$ for the remaining two-planet systems from the super-Earths-cold-Jupiter sample in Section \ref{sec:SECJ} with $a_{1,0} = 0.5\,$au and $K = 5$ for all planets}. The left and right vertical dashed lines show the 95\% instability and stability boundaries, respectively. The variables are defined in Equations (\ref{eq:sta_boundary}) and (\ref{eq:Y}).
    \label{fig:SEJ_stability}
\end{figure}

When the boundary in Equation (\ref{eq:sta_boundary}) is satisfied, a planetary system tends to be stable \citep{Petrovich2015}. More precisely, when $r_\mathrm{ap} < Y - 0.55$, there is a $>$ 95\% probability that the two-planet system is long-term unstable; when $r_\mathrm{ap} > Y + 0.25$, there is a $>$ 95\% probability that the two-planet system is long-term stable \citep{Petrovich2015}. Figure \ref{fig:SEJ_stability} shows $a_\mathrm{out} / a_\mathrm{in}$ vs.\ $r_\mathrm{ap} - Y$ of the remaining two-planet systems from the super-Earths-cold-Jupiter sample in Section \ref{sec:SECJ}. The two dashed lines are the above $95\%$ probability criteria. 91\% of the two-planet systems have more than 95\% probability that they will remain stable. In addition, all systems satisfy the stability boundary in Equation (\ref{eq:sta_boundary}). By a conservative estimation, more than 86\% of the surviving two-planet systems are stable. The proportion of stable systems could be even higher if we consider the effect of GR precession \citep{Anderson2020}.

\section{Conclusions}\label{sec:conclusion}
In this work, we simulated multi-planet systems using the $N$-body integration package \texttt{REBOUND} to study their evolution and the planets ejected from them. We analyzed the effects of various parameters on the evolution of planetary systems and how they become unstable due to collisions or ejections. Throughout our analysis, we measure time in units of the orbital period $P_1$ of the innermost planet in the system.

In our super-Earths only samples, for the same orbital spacing $K$, defined in Equation (\ref{eq:separation}), systems farther away from their host stars experience the first planet loss by ejection or collision later in their evolution, though it takes the same amount of time for them to have the first close encounters. This demonstrates that complex dynamical processes exist between the first encounters and the first planet loss events. For the systems with a large semimajor axis of the innermost orbit $a_{1,0}$, we find that the half-loss time $\tilde{t}_{1/2}$ is almost independent of $K$, with $\log \tilde{t}_{1/2}$ increasing linearly with $\log a_{1,0}$. However, for the systems with smaller $a_{1,0}$, $\tilde{t}_{1/2}$ increases with $K$. The latter can be explained by the rapid increase of the first close encounter time $\tilde{t}_\mathrm{FCE}$ with increasing $K$. In addition, we explore the dependence of the number of ejections on the Safronov number in equal-mass planetary systems. We show that the average numbers of ejections per planet are similar in systems of different planet masses but with the same Safronov number. When the Safronov number increases, the number of ejections increases as well.

In the evolution of the super-Earths-cold-Jupiter systems, orbits of super-Earth move inwards or outwards under the interaction with other planets. When the orbit of a super-Earth intersects with that of the cold Jupiter, it becomes unstable and the super-Earth will eventually be lost through collision or ejection. Because of the large Safronov number of the cold Jupiters, 38\% of super-Earths escape in the super-Earths-cold-Jupiter sample. They have a low velocity relative to their host stars, and thus, their observed velocity distribution in the Milky Way will be similar to that of their host stars. Since the cold Jupiters dominate the super-Earths-cold-Jupiter systems, samples of different $K$'s have the same fraction of ejected planets. Most super-Earths in the surviving two-planet systems have migrated inwards, and they have considerably greater eccentricities than their initial states. Under conservative estimation without GR effects, more than 86\% of the remaining two-planet systems are long-term stable, according to the criteria of \citet{Petrovich2015}. It should be noted that we do not see any binary objects ejected from our samples by checking the time interval between ejections, and so it has little impact on producing systems such as JuMBOs \citep{Pearson2023}.

In our simulations, we assumed perfect merger (i.e., two colliding planets always merge to form a planet with the sum of the masses). For the simulations with equal-mass planets, close encounters between the planets typically increase the relative velocity to less than the escape velocity before a collision, and perfect merger is a reasonable approximation \citep[e.g.,][]{Genda2012, Stewart2012}. For the super-Earth-cold-Jupiter simulations, a super-Earth could gain random velocity up the surface escape velocity of Jupiter in an encounter with the cold Jupiter, which could lead to relative velocity in a subsequent collision with another super-Earth that exceeds their surface escape velocities. The effects of a more realistic collision algorithm that accounts for other collision outcomes such as hit-and-run will require further investigation.

To further study the properties of ejected planets from multi-planet systems, we need to make progress from both theory and observations. First, future theoretical works need to adopt more realistic initial conditions, including more planets and/or the presence of protoplanetary disks in the simulations. Second, it will be vitally important to discover more FFPs with masses to improve the current statistical results (e.g., \citet{Sumi2023}), which is indeed one of the science drivers of upcoming space missions such as Roman \citep{Penny2019} and ET \citep{GeJian2022}.

\begin{acknowledgments}
    R. Z. acknowledges support from a Paul M. Doty Distinguished Graduate Fellowship and a University Graduate Fellowship. This work is partly supported by the National Science Foundation of China (Grant No.\ 12133005) and the Research Grants Council of Hong Kong (Grant No.\ 17306720). The authors acknowledge the Tsinghua Astrophysics High-Performance Computing platform at Tsinghua University for providing computational and data storage resources that have contributed to the research results reported within this paper.
\end{acknowledgments}

\bibliographystyle{aasjournal.bst}
\bibliography{Zhai}

\end{document}